\documentclass[a4paper,12pt]{article}
\pdfoutput=1 
\usepackage{jheppub} 
\usepackage{epsf}
\usepackage{amsmath,amssymb}
\usepackage[normalem]{ulem}
\input{colordvi.tex}
\usepackage[usenames,dvipsnames]{color}
\usepackage{graphicx}
\usepackage{ascmac}
\usepackage{hyperref}
\usepackage{slashed}
\usepackage{tikz-feynman}
\tikzfeynmanset{compat=1.0.0}

\newcommand{\Mpl}{M_{\rm Pl}}
\newcommand{\Lag}{\mathcal{L}}

\usepackage{hyperref} 
\hypersetup{
    colorlinks=true,       
    linkcolor=magenta,          
    citecolor=blue,        
    filecolor=magenta,      
    urlcolor=blue           
}

\begin{document}

\begin{flushright}
TU-1313,
KEK-QUP-2026-0011
\end{flushright}

\title{Inflation and Reheating by Dynamical Torsion}

\author[a]{Xiaolin Ma,}
\author[b]{Kamil Mudrunka,}
\author[a,b]{Kazunori Nakayama}

\affiliation[a]{International Center for Quantum-field Measurement Systems for Studies of the Universe and Particles (QUP), KEK, Tsukuba, Ibaraki 305-0801, Japan}
\affiliation[b]{Department of Physics, Tohoku University, Sendai, Miyagi 980-8578, Japan}

\emailAdd{xlmphy@post.kek.jp}
\emailAdd{mudrunka.kamil.q5@dc.tohoku.ac.jp}
\emailAdd{kazunori.nakayama.d3@tohoku.ac.jp}

\abstract{
In gravity theories with torsion, inflation can be driven by the dynamical torsion. Since it naturally has a derivative coupling to the axial current consisting of Standard Model particles, the reheating dynamics should be much different from conventional inflation models like Starobinsky inflation. We calculate the inflaton decay rate in detail and find that, although the 2 body decay into the fermion pair is vanishing in the massless fermion limit, the anomaly-induced 2 body decay into the gauge bosons and 3 body decay involving the Higgs boson are non-vanishing and they give sizable contributions to the total decay width of the inflaton.
In the minimal reheating scenario, the illustrative requirement of spectral index $n_s\gtrsim0.960$ gives the conditional lower bound on the reheating temperature $T_{\rm RH}\gtrsim10^5\,{\rm GeV}$.}

\maketitle
\flushbottom



\section{Introduction}

Since the idea of inflation appeared~\cite{Guth:1980zm,Sato:1981qmu,Kazanas:1980tx,Linde:1981mu,Albrecht:1982wi}, there are continuous efforts to identify the inflaton as a dynamical scalar degree of freedom in the gravity sector, starting from the Starobinsky's original model~\cite{Starobinsky:1980te}.
There have been huge developments in theoretical understandings of physics of inflation and the density perturbation, and precise observations of cosmic microwave background give stringent constraints on inflation models.
Interestingly, still the Starobinsky inflation remains as a viable model.
This might indicate that the inflaton is somehow related to the gravity sector, and hence it may be worth studying how the inflaton can be embedded in more general formulations of gravity.

We consider the so-called metric-Affine~\cite{Hehl:1994ue,Iosifidis:2019dua,Baldazzi:2021kaf} or Einstein-Cartan~\cite{Hehl:1976kj,Shapiro:2001rz} formulation of gravity.
In both formulations, we do not impose the torsion-free condition, in contrast to the Einstein's general relativity.
In the metric-Affine gravity, we further abandon the metricity condition, $\nabla_\rho g_{\mu\nu}=0$.
Therefore the torsion or non-metricity degrees of freedom appear in these formulations (which are collectively called ``distortion'' in this paper), and they may be responsible for inflation.
Studies in this direction are found in Refs.~\cite{Aoki:2020zqm,Pradisi:2022nmh,Salvio:2022suk,Gialamas:2022xtt,DiMarco:2023ncs,He:2024wqv,He:2025bli,He:2025fij,Katsoulas:2025srh,Racioppi:2025igu,Gialamas:2026pjo,DiBenedetto:2026onj,DelGrosso:2026zbg}, and actually it is found that viable models can be constructed.

In the most models in which torsions are identified as the inflaton, the inflaton originates from the parity-odd Holst curvature invariant $E^{\mu\nu\rho\sigma}R_{\mu\nu\rho\sigma}$ and is called a pseudoscalaron. It has derivative couplings to the other particles including Standard Model particles.
Due to its derivative-coupling nature, the decay pattern of the inflaton and the reheating dynamics are nontrivial.
The reheating in such a model is discussed in Refs.~\cite{Salvio:2022suk,DiBenedetto:2026onj}.
In contrast to the Starobinsky inflation, there are no 2 body decays into massless particles in a minimal inflaton coupling scheme.
Ref.~\cite{Salvio:2022suk} introduced a non-minimal coupling to the Higgs, while Ref.~\cite{DiBenedetto:2026onj} introduced an inflaton coupling to the right-handed neutrinos for successful reheating and baryogenesis.
Ref.~\cite{He:2025fij} investigated torsion couplings and possible roles for reheating in general.

In this paper we revisit the inflaton decay and reheating in models with inflation with torsion.
We will see that, even if we only introduce the minimal derivative couplings of the torsion with Standard Model (massless) fermions, it can decay through several processes.
One is the 3 body decay involving fermion pair plus the Higgs boson, and the other is the anomaly-induced 2 body decay into the gauge bosons.
They give sizable contributions to the total inflaton decay width, and hence the reheating is successful without introducing any non-minimal couplings.
Taking $n_s\gtrsim0.960$ as an illustrative requirement yields the benchmark $T_{\rm RH}\gtrsim10^5\,{\rm GeV}$ in this minimal scenario. Lower temperatures are also compatible with the measured scalar tilt, as discussed in Sec.~\ref{sec:obs}.

This paper is organized as follows.
In Sec.~\ref{sec:basics} we briefly review some basics of gravity theories with torsion.
In Sec.~\ref{sec:inf} we show how the torsion can become dynamical and inflation happens.
In Sec.~\ref{sec:reheat} we study how the inflaton decays including the effect of Yukawa interactions as well as anomaly and estimate the reheating temperature.
In Sec.~\ref{sec:obs} several observational implications are discussed.
Sec.~\ref{sec:conc} is devoted to conclusions and discussion.

\section{Basics of gravity theories with torsion}
\label{sec:basics}

In the most general formulation of gravity, we start from the general Affine connection $\Gamma^\rho_{~\mu\nu}$. We define the torsion tensor as follows:
\begin{align}
	{T^\rho}_{\mu\nu} \equiv {\Gamma^\rho}_{\mu\nu} - {\Gamma^\rho}_{\nu\mu}.
    \label{torsion}
\end{align}
We also introduce the non-metricity tensor as
\begin{align}
	Q_{\rho\mu\nu} \equiv -\nabla_{\rho} g_{\mu\nu},
    \label{nonmetricity}
\end{align}
where the covariant derivative is defined by
\begin{align}
	&\nabla_\rho V^{\mu_1\mu_2\cdots}_{~~~\nu_1\nu_2\cdots} = \partial_\rho V^{\mu_1\mu_2\cdots}_{~~~\nu_1\nu_2\cdots}
	+\sum_i \Gamma^{\mu_i}_{~\rho\lambda} V^{\cdots \mu_{i-1} \lambda \mu_{i+1}\cdots}_{~~~~~\nu_1\nu_2\cdots} 
	-\sum_i \Gamma^{\lambda}_{~\rho\nu_i} V^{\mu_1\mu_2\cdots}_{~\cdots  \nu_{i-1} \lambda \nu_{i+1} \cdots}.
\end{align}
In the conventional Einstein gravity, both torsion and non-metricity are taken to be zero. In the Einstein-Cartan formulation, the non-metricity is taken to be zero while the torsion is kept, while in the metric-Affine formulation both torsion and non-metricity are kept.
Below we basically follow the metric-Affine formulation in order to be as general as possible.

The torsion and non-metricity both behave as tensors with correct transformation properties, and they can be dynamical fields in principle, depending on the action we will consider.
From the definitions (\ref{torsion}) and (\ref{nonmetricity}), the Affine connection can be written as
\begin{align}
	{\Gamma^\rho}_{\mu\nu} = \overset{\circ}{\Gamma}\,^\rho_{~\mu\nu} + {K^\rho}_{\mu\nu}, \label{Affine}
\end{align}
where $\overset{\circ}{\Gamma}\,^\rho_{~\mu\nu}$ denotes the Levi-Civita connection: $\overset{\circ}{\Gamma}\,^\rho_{~\mu\nu} \equiv \frac{1}{2}g^{\rho\sigma}\left( \partial_\mu g_{\nu\sigma} + \partial_\nu g_{\sigma\mu}-\partial_{\sigma}g_{\mu\nu} \right)$ and ${K^\rho}_{\mu\nu}$ is called distortion tensor. It is expressed in terms of the torsion and non-metricity as
\begin{align}
    {K^\rho}_{\mu\nu} = \frac{1}{2}\left( {T^\rho}_{\mu\nu} + T_{\nu~\mu}^{~\rho} -T_{\mu\nu}^{~~\rho}  +Q_{\mu\nu}^{~~\rho} + Q_{\nu~\mu}^{~\rho}  -Q_{~\mu\nu}^{\rho}\right).
    \nonumber
\end{align}
The Riemann tensor is then given by
\begin{align}
	R^{\alpha}_{~\beta\mu\nu}
    &=\partial_\mu\Gamma^\alpha_{~\nu\beta} - \partial_\nu\Gamma^\alpha_{~\mu\beta}  
	+\Gamma^\alpha_{~\mu\lambda}\Gamma^\lambda_{~\nu\beta} -\Gamma^\alpha_{~\nu\lambda}\Gamma^\lambda_{~\mu\beta}\\
    &= \overset{\circ}{R}\,^{\alpha}_{~\beta\mu\nu}
	+\overset{\circ}\nabla_\mu K^\alpha_{~\nu\beta} - \overset{\circ}\nabla_\nu K^\alpha_{~\mu\beta}
	+K^\alpha_{~\mu\lambda}K^\lambda_{~\nu\beta} - K^\alpha_{~\nu\lambda}K^\lambda_{~\mu\beta},
	\label{R_gK}
\end{align}
where $\overset{\circ}{R}\,^{\alpha}_{~\beta\mu\nu}$ denotes the Riemann tensor constructed only from the metric, and $\overset{\circ}{\nabla}_\mu$ is the covariant derivative with respect to the Levi-Civita connection, both of which are the same quantities as those in the usual general relativity.
The Ricci scalar is given by
\begin{align}
	R= \overset{\circ}{R}
	+\overset{\circ}\nabla_\mu K^{\mu\nu}_{~~\nu} - \overset{\circ}\nabla_\nu K^{~\mu\nu}_{\mu}
	+K^\mu_{~\mu\lambda}K^{\lambda\nu}_{~~\nu} - K^{\mu\nu}_{~~\lambda}K^\lambda_{~\mu\nu}.
    \label{Ricci_K}
\end{align}

By construction, the torsion ${T^\rho}_{\mu\nu}$ is antisymmetric under $\mu\leftrightarrow \nu$, while the non-metricity $Q_{\rho\mu\nu}$ is symmetric under $\mu\leftrightarrow \nu$. 
Having this in mind, we can conveniently decompose these tensors as follows:
\begin{align}
	&\hat T^\mu \equiv E^{\mu\nu\rho\sigma} T_{\nu\rho\sigma}, \label{Tmu}\\
	&T^\mu \equiv g_{\rho\sigma} T^{\rho\mu\sigma}, \\
	&\widetilde T_{\rho\mu\nu} \equiv T_{\rho\mu\nu}+\frac{1}{3}(g_{\rho\mu}T_\nu-g_{\rho\nu}T_{\mu})-\frac{1}{6}E_{\rho\mu\nu\sigma}\hat T^\sigma,
\end{align}
where $E^{\mu\nu\rho\sigma}$ is the Levi-Civita tensor, $E^{\mu\nu\rho\sigma}=\epsilon^{\mu\nu\rho\sigma}/\sqrt{-g}$ and $E_{\mu\nu\rho\sigma}=-\sqrt{-g}\,\epsilon_{\mu\nu\rho\sigma}$ with $\epsilon^{0123}=1$, and
\begin{align}
	& Q^\mu \equiv g_{\rho\sigma}Q^{\mu\rho\sigma},\qquad \hat Q^\mu \equiv g_{\rho\sigma}Q^{\rho\mu\sigma},\\
	&\widetilde Q_{\rho\mu\nu}\equiv Q_{\rho\mu\nu}-\frac{1}{18}\left[g_{\mu\nu}(5Q_\rho-2\hat Q_\rho)
    + 4(g_{\rho\mu}\hat Q_\nu +g_{\rho\nu}\hat Q_\mu) - (g_{\rho\mu}Q_\nu +g_{\rho\nu}Q_\mu) \right].
    \label{Qmunu}
\end{align}
Note that $\widetilde T_{\rho\mu\nu}$ satisfies $\epsilon^{\lambda\rho\mu\nu} \widetilde T_{\rho\mu\nu}=0$ and $g_{\rho\nu} \widetilde T^{\rho\mu\nu} = 0$, and $\widetilde Q^{\rho\mu\nu}$ satisfies $g_{\mu\nu}\widetilde Q^{\rho\mu\nu}=0$ and $g_{\rho\nu}\widetilde Q^{\rho\mu\nu}=0$.
Substituting these decompositions into (\ref{Ricci_K}), we obtain~\cite{Rigouzzo:2023sbb,Nakayama:2026nhg}
\begin{align}
	R =& \overset{\circ}{R} + \overset{\circ}{\nabla}_\mu(2T^\mu - Q^\mu + \hat Q^\mu) \nonumber \\
    &+\frac{1}{24}\hat T_\mu \hat T^\mu-\frac{2}{3}T_\mu(T^\mu-Q^\mu + \hat Q^\mu) 
    -\frac{11}{72}Q_\mu Q^\mu + \frac{2}{9}Q_\mu \hat Q^\mu + \frac{1}{18} \hat Q_\mu \hat Q^\mu \nonumber \\
    &+\frac{1}{2}\widetilde T_{\mu\nu\rho}\widetilde T^{\mu\nu\rho} - \widetilde T_{\mu\nu\rho} \widetilde Q^{\nu\rho\mu} + \frac{1}{4}\widetilde Q_{\mu\nu\rho}(\widetilde Q^{\mu\nu\rho}-2 \widetilde Q^{\rho\mu\nu}).
    \label{Ricci_original}
\end{align}
Vector components $T_\mu, Q_\mu$ and $\hat Q_\mu$ have quadratic terms and it is useful to go to the diagonal basis. The corresponding eigenstates are given by~\cite{Nakayama:2026nhg}
\begin{align}
	N^0_\mu = \frac{1}{\sqrt{77}}\left(3T_\mu +8 Q_\mu + 2\hat Q_\mu \right),
    \label{Nmu}
\end{align}
and
\begin{align}
	N_\mu^{\pm} = D_\pm\left( \frac{59\pm\sqrt{6721}}{54}T_\mu - \frac{95\pm\sqrt{6721}}{144}Q_\mu + \hat Q_\mu \right),
\end{align}
where $D_\pm$ is a normalization constant.
By using these vectors, we can rewrite the Ricci scalar as
\begin{align}
	R =& \overset{\circ}{R} + \overset{\circ}{\nabla}_\mu( c_+ N_\mu^+ + c_- N_\mu^- )
    +\frac{1}{24}\hat T_\mu \hat T^\mu  - \sum_{i=\pm} \frac{M_i^2}{M_{\rm Pl}^2}N^i_\mu N_i^\mu \nonumber \\
    &+\frac{1}{2}\widetilde T_{\mu\nu\rho}\widetilde T^{\mu\nu\rho} - \widetilde T_{\mu\nu\rho} \widetilde Q^{\nu\rho\mu} + \frac{1}{4}\widetilde Q_{\mu\nu\rho}(\widetilde Q^{\mu\nu\rho}-2 \widetilde Q^{\rho\mu\nu}),
    \label{Ricci}
\end{align}
where
\begin{align}
	M_\pm^2= \frac{55\pm \sqrt{6721}}{144}\times M_{\rm Pl}^2,
    \label{mass_eigen}
\end{align}
with $M_{\rm Pl}$ being the reduced Planck scale and we have defined $c_\pm D_\pm=\frac{1}{2}\left(1 \pm \frac{49}{\sqrt{6721}} \right)$ so that $2T_\mu-Q_\mu +\hat Q_\mu = c_+ N_\mu^+ + c_- N_\mu^-$.
Note that $N_\mu^0$ does not appear at all in $R$. This is because $R$ (as well as $\hat R$ defined below) is invariant under the projective transformation $\Gamma^\rho_{~\mu\nu}\to\Gamma^\rho_{~\mu\nu}+\delta^\rho_{~\nu}\xi_\mu$ with arbitrary vector $\xi_\mu$, while $N_\mu^0$ is not~\cite{Sandberg:1975db,Hehl:1978zkk}. 

In theories with torsion, the following so-called Holst term~\cite{Hojman:1980kv,Nelson:1980ph,Holst:1995pc} is nonzero:
\begin{align}
	\hat R &= E^{\mu\nu\rho\sigma} R_{\mu\nu\rho\sigma} \nonumber\\
    &= - \overset{\circ}{\nabla}_\mu \hat T^\mu-\frac{1}{3}\hat T^\mu(2T_\mu-Q_\mu + \hat Q_\mu) + E^{\mu\nu\rho\sigma}\left(\frac{1}{2}\widetilde T_{\lambda\mu\nu} - \widetilde Q_{\mu\nu\lambda}\right)\widetilde T^{\lambda}_{~\rho\sigma}.
    \label{Holst}
\end{align}
The Ricci scalar (\ref{Ricci}) and Holst term (\ref{Holst}) are basic ingredients of theories discussed in the following sections.
If one wants to consider the Einstein-Cartan theory, one can just take $Q_\mu=\hat Q_\mu = \widetilde Q_{\mu\nu\rho}=0$ in Eqs.~(\ref{Ricci_original}) and (\ref{Holst}).

\section{Inflation from dynamical torsion}
\label{sec:inf}

Now we briefly review how inflation successfully happens by identifying distortion fields as the inflaton.
Let us consider an action as
\begin{align}
	S=\int d^4x \sqrt{-g}\left[ \frac{M_{\rm Pl}^2}{2} R + f(\hat R) \right],
	\label{action_fR}
\end{align}
where $f(\hat R)$ is some function of $\hat R$, which is not specified here.\footnote{
    We here assume the absence of terms quadratic in $N_{\mu}^\pm$ that cannot be forbidden by the projective invariance. Ref.~\cite{Barker:2024dhb} introduced an extended projective symmetry to forbid some of such terms.
} Introducing an auxiliary field $\chi$, this action is written as
\begin{align}
	S=\int d^4x \sqrt{-g}\left[ \frac{M_{\rm Pl}^2}{2} R + F(\chi) \hat R + f(\chi) - \chi F(\chi)\right],
	\label{S_fchi}
\end{align}
where $F(\chi) \equiv \frac{\partial f}{\partial \chi}$. By solving the equation of motion of $\chi$, we obtain $\chi=\hat R$ and it reduces to the action (\ref{action_fR}) as far as $\frac{\partial F}{\partial\chi} \neq 0$.

\subsection{Metric-Affine gravity}

Let us substitute (\ref{Ricci}) and (\ref{Holst}) into (\ref{S_fchi}). We soon notice that it is only $\hat T_\mu$ that has a derivative term and all others are non-dynamical.
Pure tensor parts are found to be zero by solving their equations of motion: $\widetilde T_{\mu\nu\rho} = \widetilde Q_{\mu\nu\rho} = 0$.
For vector degrees, let us rewrite them in terms of the mass eigenstate:
\begin{align}
	S=&\int d^4x \sqrt{-g}\left[ \frac{M_{\rm Pl}^2}{2}\overset{\circ}{R} + \frac{M_{\rm Pl}^2}{48}\hat T_\mu \hat T^\mu
	-\sum_{i=\pm}\frac{1}{2} M_i^2 N_{i,\mu} N_i^\mu \right. \nonumber\\
	&\left. + (\partial_\mu F)\,\hat T^\mu -\frac{F(\chi)}{3}\hat T_\mu(c_+ N_+^\mu + c_- N_-^\mu) + f(\chi) - \chi F(\chi) \right].
\end{align}
Note that, since both $R$ and $\hat R$ are projective invariant, the ``massless'' state (\ref{Nmu}) does not appear in the action.
By solving the equation of motion of $N_\pm^\mu$, we find
\begin{align}
	N^\mu_\pm = -\frac{c_\pm F(\chi)}{3M_\pm^2} \hat T^\mu.
\end{align}
From the equation of motion of $\hat T_\mu$, we find
\begin{align}
	\hat T_\mu = -\frac{24}{M_{\rm Pl}^2} \frac{\partial_\mu F}{1+\frac{8F^2(\chi)}{3M_{\rm Pl}^2}\left( \frac{c_+^2}{M_+^2} + \frac{c_-^2}{M_-^2} \right) } =  -\frac{24}{M_{\rm Pl}^2} \frac{\mathcal F(\chi)}{1+\frac{16F^2(\chi)}{M_{\rm Pl}^4}} \partial_\mu\chi,
	\label{T_chi}
\end{align}
where $\mathcal F(\chi) \equiv \frac{\partial F}{\partial \chi} = \frac{\partial^2 f}{\partial\chi^2}$.
Substituting this back into the action, we obtain
\begin{align}
	S=&\int d^4x \sqrt{-g}\left[ \frac{M_{\rm Pl}^2}{2}\overset{\circ}{R} 
		-\frac{12}{M_{\rm Pl}^2} \frac{\mathcal F^2(\chi)}{1+\frac{16F^2(\chi)}{M_{\rm Pl}^4}} (\partial_\mu \chi)^2 - V(\chi) \right],
    \label{Schi_metricAffine}
\end{align}
where $V(\chi) = \chi F(\chi) - f(\chi)$.
Thus it reduced the action of Einstein gravity plus a dynamical real scalar with potential $V(\chi)$. The canonically normalized field is given by
\begin{align}
	\phi = \frac{1}{M_{\rm Pl}} \int \sqrt{\frac{24 \mathcal F^2(\chi)}{1+ \frac{16 F^2(\chi)}{M_{\rm Pl}^4}}} \, d\chi.
    \label{phi_chi}
\end{align}
Looking at Eq.~(\ref{T_chi}), it is clear that it is a scalar component of the torsion vector $\hat T_\mu$ that becomes dynamical.

\begin{figure}[t]
\centering
\includegraphics[width=0.6\textwidth]{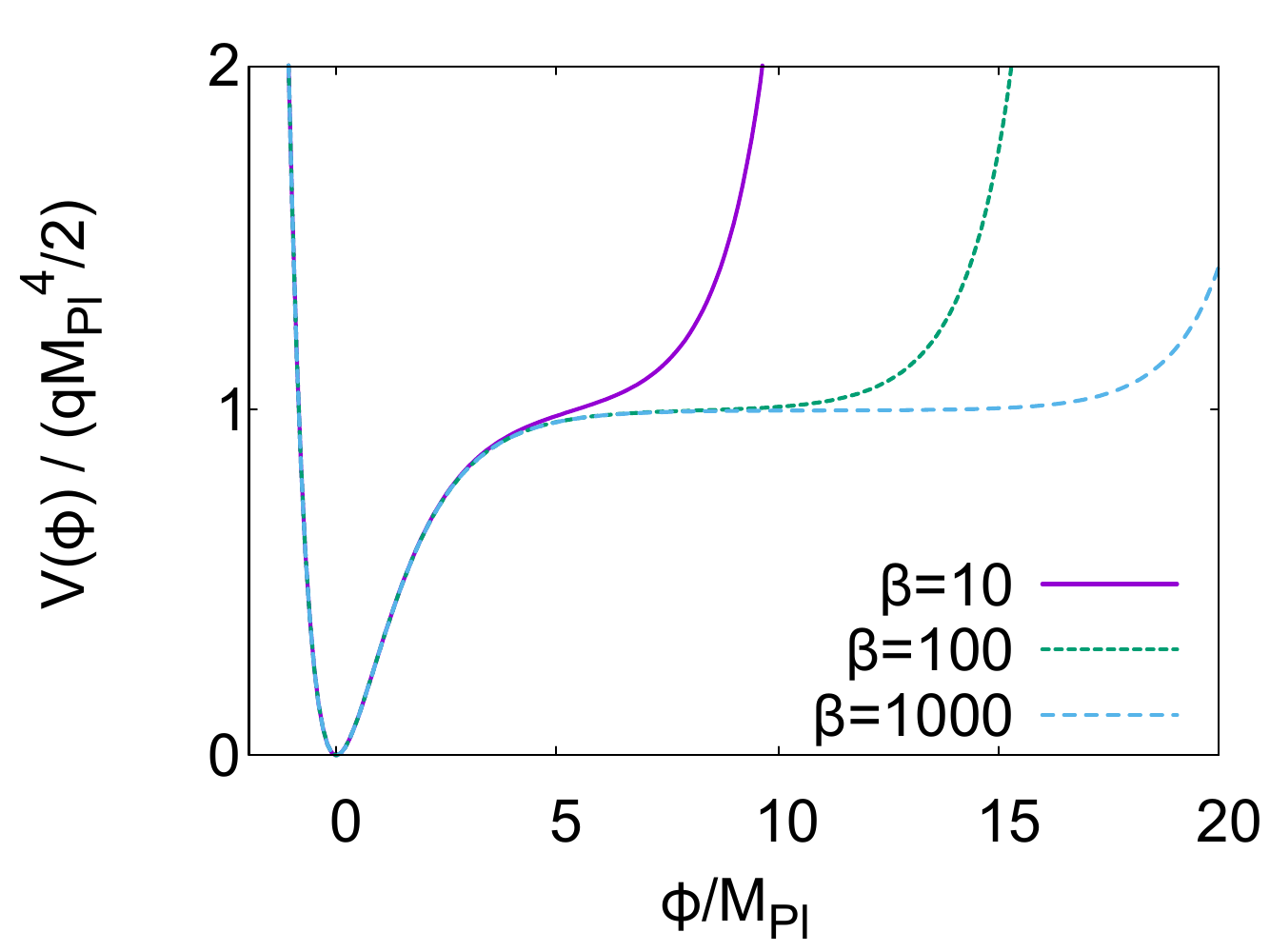}
\caption{Scalar potential (\ref{eq:app-exact-canonical-potential}) for the canonical inflaton $\phi$ for several choice of $\beta$.}
\label{fig:V_holst}
\end{figure}

Eq.~(\ref{phi_chi}) tells us that we need at least quadratic term in the function $f(\chi)$ in order to have a dynamical degree of freedom.
Below we consider the simplest possible function for $f(\chi)$:
\begin{align}
	f(\chi) = -\beta M_{\rm Pl}^2\,\chi + \frac{\gamma}{2}\chi^2,
\end{align}
where $\beta$ and $\gamma$ are dimensionless positive constants.\footnote{
    The parameter $-(4\beta)^{-1}$ is called the Barbero-Immirzi parameter~\cite{BarberoG:1994eia,Immirzi:1996di}.
} Then the potential is simply given by
\begin{align}
	V(\chi) = \frac{\gamma}{2}\chi^2.
\end{align}
The relation between $\chi$ and the canonical field $\phi$ is obtained from Eq.~(\ref{phi_chi}) as
\begin{equation}
\chi=\frac{\Mpl^2}{\gamma}
\left(\beta+\frac{1}{4}\sinh x\right),~~~~~~x\equiv\sqrt{\frac{2}{3}}\frac{\phi}{\Mpl}
-\operatorname{arsinh}(4\beta),
\label{eq:app-exact-x-definition}
\end{equation}
The potential is then expressed as
\begin{align}
V(\phi)&=\frac{q\Mpl^4}{2}
\left(1+\frac{\sinh x}{4\beta}\right)^2,~~~~~~q\equiv\frac{\beta^2}{\gamma}.
\label{eq:app-exact-canonical-potential}
\end{align}
The potential minimum is at $\phi=0$, where $x=-\operatorname{arsinh}(4\beta)$.
The inflaton has a quadratic potential around the minimum and its mass is given by
\begin{equation}
m_\phi^2
=\frac{q(1+16\beta^2)}{24\beta^2}\Mpl^2.
\label{eq:inflaton-mass}
\end{equation}
The scalar potential (\ref{eq:app-exact-canonical-potential}) is plotted in Fig.~\ref{fig:V_holst}.
{For $\gamma>0$, eliminating the connection gives Einstein gravity coupled to a canonical scalar with a nonnegative potential and positive vacuum mass. Below $m_\phi$, integrating out the scalar leaves General Relativity with suppressed higher-dimensional interactions.}

For large $\beta$ $(\gg 1)$, there appears a long plateau region in the potential.
For $\beta\gg 1$, the relation between $\phi$ and $\chi$ are given by
\begin{align}
	\phi \simeq -\sqrt{\frac{3}{2}}M_{\rm Pl}\log\left(1-\frac{\gamma\chi}{\beta M_{\rm Pl}^2} \right)
	~~~\leftrightarrow~~~
	\chi \simeq \frac{\beta M_{\rm Pl}^2}{\gamma} \left[ 1-\exp\left(-\sqrt{\frac{2}{3}}\frac{\phi}{M_{\rm Pl}}\right) \right],
\end{align}
for $\chi \lesssim \chi_* =\frac{\beta M_{\rm Pl}^2}{\gamma}\left(1-\frac{1}{4\beta}\right)$ or $\frac{\phi}{M_{\rm Pl}} \lesssim \sqrt{\frac{3}{2}}\log(4\beta)$. In this range, the scalar potential is approximated as
\begin{equation}
V(\phi)\simeq\frac{q\Mpl^4}{2}
\left[1-\exp\left(-\sqrt{\frac{2}{3}}
\frac{\phi}{\Mpl}\right)\right]^2,
\label{eq:app-starobinsky-limit}
\end{equation}
which is the same as the Starobinsky inflation potential~\cite{Starobinsky:1980te}.
This class of inflation models in either metric-Affine or Einstein-Cartan theories have been extensively studied in Refs.~\cite{Pradisi:2022nmh,Salvio:2022suk,He:2024wqv,Katsoulas:2025srh,DiBenedetto:2026onj,DelGrosso:2026zbg}.

Therefore, in the large $\beta$ limit, the inflationary dynamics becomes the same as the Starobinsky inflation.\footnote{
    Here we note that $f(R)$ theories, including the Starobinsky-type model, do not have a scalaron degree of freedom in the metric-Affine or Einstein-Cartan gravity, or, more generally, in the Palatini formalism of gravity~\cite{Sotiriou:2006qn,Sotiriou:2008rp,Olmo:2011uz}. This is most easily shown when working with $(g_{\mu\nu},\Gamma^\rho_{~\mu\nu})$ as independent variables, while it is a bit nontrivial practice when working with $(g_{\mu\nu},K^\rho_{~\mu\nu})$~\cite{He:2024wqv}.
}
However, as we will see below, the way inflaton couples to the Standard Model matter fields is much different from the Starobinsky case. 
It indicates that these models are distinguishable by carefully looking at the reheating dynamics.
Thus it is important to explore details of the reheating in the present model.

\subsection{Einstein-Cartan gravity}

We can follow the similar procedure to the metric-Affine case in order to obtain the torsion action in the Einstein-Cartan gravity.
By substituting (\ref{Ricci_original}) and (\ref{Holst}) with $Q_\mu=\hat Q_\mu = \widetilde Q_{\mu\nu\rho} =0$ into (\ref{S_fchi}) and using the equation of motion of $T_\mu$, we find
\begin{equation}
T_\mu = -\frac{F(\chi)}{M_{\rm Pl}^2} \hat T_\mu.
\end{equation}
Then the equation of motion of $\hat T_\mu$ yields the same equation as (\ref{T_chi}). 
By substituting it to the action (\ref{S_fchi}), the resulting action is also the same as (\ref{Schi_metricAffine}).
{For the specified pure-gravity action, eliminating the connection gives the same effective scalar action in the Einstein--Cartan and metric-Affine formulations. }
Therefore, the analyses in the following sections are equally applied to either the Einstein-Cartan or metric-Affine theories.

\section{Reheating} \label{sec:reheat}

\subsection{Torsion-matter couplings}

To analyze the reheating, we must specify the couplings between the torsion and the Standard Model particles. 
A natural way to introduce a torsion-matter coupling is just to replace the covariant derivative $\overset{\circ}{\nabla} \to \nabla$ in the matter sector.
Then the torsion-fermion couplings naturally appear, while torsion couplings to gauge bosons do not appear due to the gauge invariance.
For torsion-scalar couplings, there are some ambiguities. See e.g. Refs.~\cite{Shimada:2018lnm,Rigouzzo:2023sbb,He:2025fij,Nakayama:2026nhg} for more details.
In this paper we mostly consider ``minimal'' torsion-fermion couplings.
{Fermions can also be consistently defined and coupled in metric-Affine gravity with nonmetricity~\cite{Rigouzzo:2023sbb}.}

To derive the torsion-fermion couplings, let us expand the torsion around its potential minimum.
Around the potential minimum, $F$ and $\mathcal F$ are approximated by constants: $F \simeq -\beta M_{\rm Pl}^2$ and $\mathcal F \simeq \gamma$. Then, from Eqs.~(\ref{T_chi}) and (\ref{phi_chi}), the torsion vector $\hat T_\mu$ is given by
\begin{align}
	\hat T_\mu \simeq - \frac{1}{M_{\rm Pl}} \sqrt{\frac{24}{1+16\beta^2}} \partial_\mu\phi.		
\end{align}
Torsion naturally couples to fermions in the kinetic term as
\begin{align}
\frac{i}{2}\overline\psi \gamma^\mu \nabla_\mu \psi + {\rm h.c.}=  \left(\frac{i}{2}\overline\psi \gamma^\mu \overset{\circ}{\nabla}_\mu \psi +{\rm h.c.}\right)
	-\frac{1}{8}\hat T_\mu\,\overline\psi\gamma_5\gamma^\mu \psi.
\end{align}
In terms of chiral fermions, it is equivalently written as
\begin{align}
\frac{i}{2}\overline\psi_{\bullet} \gamma^\mu \nabla_\mu \psi_{\bullet} + {\rm h.c.}= 
	\left(\frac{i}{2}\overline\psi_{\bullet} \gamma^\mu \overset{\circ}{\nabla}_\mu \psi_{\bullet} + {\rm h.c.}\right)
	\pm\frac{1}{8} \hat T_\mu\,\overline\psi_{\bullet}\gamma^\mu \psi_{\bullet},
\end{align}
where ${\bullet}=$ L (left chirality) or R (right chirality), and the plus (minus) sign corresponds to L (R).
Thus the canonical scalar torsion $\phi$ couples to the fermion as
\begin{align}
	\mathcal L = \frac{1}{4M_{\rm Pl}} \sqrt{\frac{6}{1+16\beta^2}} (\partial_\mu\phi)\overline\psi\gamma_5\gamma^\mu \psi
    =\frac{\partial_\mu\phi}{f_{\rm eff}}\,
\bar\psi\gamma^\mu\gamma_5\psi,
	\label{eq:canonical-torsion-coupling}
\end{align}
where
\begin{equation}
\frac{1}{f_{\rm eff}}
\equiv
\frac{1}{4\Mpl}\sqrt{\frac{6}{1+16\beta^2}} .
\label{eq:feff-definition}
\end{equation}
This is the minimal inflaton-matter coupling. The inflaton derivatively couples to the axial current consisting of all the Standard Model fermions in a natural way.
Due to this derivative coupling nature, $\phi$ is often called pseudo-scalaron, in contrast to the scalaron in the Starobinsky inflation context.

However, this interaction term does not induce $\phi$ decay into the fermion pair if fermions are massless. This is most easily seen by integrating by parts and use $\partial_\mu (\overline\psi\gamma_5\gamma^\mu \psi) = 2im_\psi \overline\psi \gamma_5\psi$ for on-shell fermions and hence it vanishes for $m_\psi\to 0$.
More concretely, the decay rate is estimated as 
\begin{equation}
    \Gamma_{\phi\to\psi\overline\psi} = \frac{m_\phi m_\psi^2}{2\pi f_{\rm eff}^2} \left(1-\frac{4m_\psi^2}{m_\phi^2}\right)^{1/2},
    \label{Gamma_phi_to_ff}
\end{equation}
if $\psi$ is a Dirac fermion, while it should be multiplied by a factor $2$ if it is a Majorana fermion.
Actually all the Standard Model fermions are effectively regarded as massless in the early universe and hence the decay rate vanishes. 
An well-motivated exception is right-handed neutrinos whose mass $(m_N)$ is relatively large, and $\phi$ may decay into them in such a case~\cite{DiBenedetto:2026onj},
although still the decay rate is suppressed by the ratio $(m_N/m_\phi)^2$.
On the other hand, Ref.~\cite{Salvio:2022suk} introduced a non-minimal torsion-Higgs coupling of the form 
\begin{equation}
\Lag \sim \hat T^\mu \partial_\mu |H|^2 \sim \frac{\partial^\mu\phi}{f_{\rm eff}}\left(\partial_\mu |H|^2\right).
\label{torsion-Higgs}
\end{equation}
{Terms like $\mathcal L\sim |H|^2R$ or $\mathcal L\sim |H|^2\hat R$ also lead to the coupling of this form.}
This induces the $\phi$ decay into the Higgs pair without any suppression: i.e., the decay rate is parametrically estimated as $\Gamma \simeq m_\phi^3 /(8\pi f_{\rm eff}^2)$.
{We note that more general torsion-matter interactions may also modify the effective scalar dynamics and matter couplings.}
However, in this paper we want to keep the minimal torsion coupling of the form (\ref{eq:canonical-torsion-coupling}) and study whether the $\phi$ decay is really suppressed or not.
Thus we do not include the torsion-Higgs coupling (\ref{torsion-Higgs}) hereafter.

Below we point out that, although 2 body decay into massless fermions are highly suppressed, there are sizable contributions to the $\phi$ decay through 3 body decay into the fermion pair plus Higgs boson, and anomaly-induced 2 body decay into the gauge boson pair.\footnote{
    Top-Yukawa-induced 3 body decay and the anomaly-induced 2 body decay of inflaton have been discussed in the context of supergravity~\cite{Endo:2006qk,Endo:2007ih,Endo:2007sz}, though underlying physics is not the same as our case.
}
Both processes give sizable contributions to the total decay rate with only mild suppression factor $\sim 10^{-2}$ compared with the canonical rate $m_\phi^3/(8\pi f_{\rm eff}^2)$. Therefore, we do not need to include non-minimal torsion-Higgs coupling or a right-handed neutrino (with its mass happening to be close to the torsion) in order to have successful reheating.

\subsection{Decay rate and reheating temperature} \label{sec:decay}

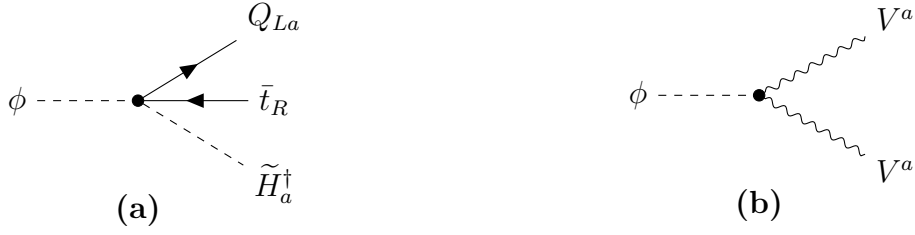
\begin{figure}[t]
\centering
\begin{minipage}[c]{0.46\textwidth}
\centering
\begin{tikzpicture}[baseline=(current bounding box.center)]
\begin{feynman}
\vertex (phi) at (-1.6,0) {$\phi$};
\vertex[dot] (v) at (0,0) {};
\vertex (q) at (1.8,1.1) {$Q_{La}$};
\vertex (tb) at (1.8,0) {$\bar t_R$};
\vertex (h) at (1.8,-1.1) {$\widetilde H_a^\dagger$};
\diagram*{
  (phi) -- [scalar] (v),
  (v) -- [fermion] (q),
  (tb) -- [fermion] (v),
  (v) -- [scalar] (h)
};
\node at (0,-1.45) {\textbf{(a)}};
\end{feynman}
\end{tikzpicture}
\end{minipage}\hfill
\begin{minipage}[c]{0.46\textwidth}
\centering
\begin{tikzpicture}[baseline=(current bounding box.center)]
\begin{feynman}
\vertex (phi) at (-1.6,0) {$\phi$};
\vertex[dot] (v) at (0,0) {};
\vertex (v1) at (1.8,1.0) {$V^a$};
\vertex (v2) at (1.8,-1.0) {$V^a$};
\diagram*{
  (phi) -- [scalar] (v),
  (v) -- [boson] (v1),
  (v) -- [boson] (v2)
};
\node at (0,-1.45) {\textbf{(b)}};
\end{feynman}
\end{tikzpicture}
\end{minipage}
\caption{Local vertices in the current-divergence basis.  Panel (a) is the
Yukawa contact interaction with $\widetilde Y_f=i(Y_fc_f-c_FY_f)$; for the
universal torsion current, $\widetilde Y_f=2iY_f$.  Panel (b) is the anomaly
operator proportional to $\alpha_a C_a/(4\pi f)$.}
\label{fig:app-equivalent-vertices}
\end{figure}

The universal coupling to Standard Model fermion particles~\eqref{eq:canonical-torsion-coupling} leads to the decay of the torsion particle into Standard Model particles. Even though the derivative interaction~\eqref{eq:canonical-torsion-coupling} gives
no on-shell two-body decay into massless fermions in the early Universe, the quantum anomaly of the chiral current plus the Yukawa interaction which explicitly breaks the chiral symmetry gives rise to the decay of the torsion inflaton into Standard Model particles.  The divergence of the current receives a classical Yukawa
contribution and a quantum gauge-anomaly contribution. We consider the dominant Yukawa contribution of Top quarks hereafter. Integrating by parts 
therefore gives the equivalent decay interaction (see Appendix~\ref{app:torsion_coupling} for detailed derivation)
\begin{align}
\Lag_{\rm decay}
={}&-\frac{2i\phi}{f_{\rm eff}}
\left({\cal O}_Y-{\cal O}_Y^\dagger\right)
\notag\\
&+\frac{\phi}{f_{\rm eff}}\left[
\frac{\alpha_s C_3}{4\pi}G_{\mu\nu}^A\widetilde G^{A\mu\nu}
+\frac{\alpha_2 C_2}{4\pi}W_{\mu\nu}^I\widetilde W^{I\mu\nu}
+\frac{\alpha_1 C_1}{4\pi}B_{\mu\nu}\widetilde B^{\mu\nu}
\right],
\label{eq:main-equivalent-decay-basis}
\end{align}
where
\begin{equation}
{\cal O}_Y
=\bar Q_L\widetilde H Y_u u_R
+\bar Q_LH Y_d d_R
+\bar L_LH Y_e e_R,
\qquad
\widetilde X^{\mu\nu}=\frac{1}{2}\epsilon^{\mu\nu\rho\sigma}
X_{\rho\sigma},
\end{equation}
and $\alpha_s=g_s^2/(4\pi)$, $\alpha_2=g_2^2/(4\pi)$, and
$\alpha_1=g_1^2/(4\pi)$ for the Standard Model SU(3), SU(2)$_L$ and U(1)$_Y$ gauge couplings.
Coefficients $C_1,C_2$ and $C_3$ are calculated in Appendix~\ref{app:2body_to_gauge} and given by $C_1=10$, $C_2=6$ and $C_3=6$.
Eq.~\eqref{eq:main-equivalent-decay-basis} makes the two leading classes
of decay channels manifest. The first line gives Yukawa-term-originated three-body
decays, dominated by the top Yukawa coupling. The second line gives anomaly-induced decays into two gauge bosons. 
Diagrammatic representations of these processes are shown in Fig.~\ref{fig:app-equivalent-vertices}.

\paragraph{3 body decay}

The first line of (\ref{eq:main-equivalent-decay-basis}) makes $\phi$ decay into fermion pair plus Higgs boson. The decay rate is given by
\begin{equation}
\Gamma_{\phi\to f\bar fH}= A_{f\bar f H}\frac{m_\phi^3}{f_{\rm eff}^2},~~~~~~A_{f\bar f H}= \frac{|y_t|^2}{32\pi^3}
\end{equation}
where $y_t$ denotes the top Yukawa coupling. Since the 3 body decay rate is proportional to the square of Yukawa coupling, processes  involving lighter fermions are highly suppressed.
We note that the 3 body decay involving a gauge boson (i.e., the inflaton decay into fermion pair plus gauge boson) vanishes in the massless fermion limit.
It may be understood as a result of chiral current conservation in the massless limit, but we can also explicitly show it by calculating the 3 body decay amplitude (see Appendix \ref{sec:Gauge3body}).

\paragraph{2 body decay}

The second line of (\ref{eq:main-equivalent-decay-basis}) makes $\phi$ decay into the gauge boson pair. The decay rate is given by
\begin{align}
&\Gamma_{\phi\to VV}=\left(A_{gg}+A_{WW}+A_{BB}\right) \frac{m_\phi^3}{f_{\rm eff}^2},\\
&A_{gg}=\frac{9\alpha_s^2}{2\pi^3},\qquad
A_{WW}=\frac{27\alpha_2^2}{16\pi^3},\qquad
A_{BB}=\frac{25\alpha_1^2}{16\pi^3}.
\end{align}
See Appendix~\ref{sec:loop2body} for technical details how to calculate the effective inflaton-gauge-boson interaction and the decay rate.


Combining the 2-body and 3-body decay processes, the total decay width is given by 
\begin{equation}
\Gamma_{\rm tot}= A_{\rm tot}\frac{m_\phi^3}{f_{\rm eff}^2},~~~~~~
 A_{\rm tot}=  A_{f\bar f H} + A_{gg}+A_{WW}+A_{BB}.
\label{eq:total-width-factorization}
\end{equation}
For the decay coefficients we use high energy Standard Model couplings given in Refs.~\cite{Buttazzo:2013uya,Mihaila:2012fm}. Numerically we take
\begin{equation}
y_t=0.50,\qquad \alpha_s=0.030,\qquad
g_1=0.35,\qquad g_2=0.65,
\end{equation}
or equivalently $\alpha_1=g_1^2/(4\pi)=9.75\times10^{-3}$ and $\alpha_2=g_2^2/(4\pi)=3.36\times10^{-2}$, leading to $A_{\rm tot}=4.49\times10^{-4}$.
The resulting branching ratios at fixed high-scale couplings could be calculated by
\begin{equation}
\begin{split}
{\rm Br}_{f\bar f H}=0.561,~~
{\rm Br}_{gg}=0.291,~~
{\rm Br}_{WW}=0.137,~~
{\rm Br}_{BB}=1.07\times10^{-2}.
\end{split}
\label{eq:benchmark-branching-ratios}
\end{equation}
Thus we see that the branching ratio of the 3 body decay and anomaly-induced 2 body decay are roughly comparable.
Recalling the definitions of $m_\phi$ \eqref{eq:inflaton-mass} and $f_{\rm eff}$ \eqref{eq:feff-definition}, we have a direct map from model parameters $(\beta,\gamma)$ to the total width. The corresponding reheating temperature {$T_{\rm RH}$} {is defined by 
\begin{equation}
T_{\rm RH}=\left(\frac{90}{\pi^2g_*}\right)^{1/4}
\sqrt{\Gamma_{\rm tot}\Mpl},
\label{eq:trh-total-width}
\end{equation}
}where the relativistic degrees of freedom is $g_*=106.75$ for high enough reheating temperature.
{For given $\beta,\gamma$ (or equivalently $\beta,q$), this defines $T_{\rm RH}$. }
{
Note that $T_{\rm RH}$ does not precisely correspond to the temperature at which the matter and radiation density becomes equal. We numerically solved the evolution equations of the inflaton and radiation energy density ($\rho_\phi$ and $\rho_{\rm R}$, respectively) through 
\begin{align}
\dot\rho_\phi+3H\rho_\phi&=-\Gamma_{\rm tot}\rho_\phi,~~~~~~
\dot\rho_{\rm R}+4H\rho_{\rm R}=\Gamma_{\rm tot}\rho_\phi,
\label{eq:continuous-reheating-quadratic}
\end{align}
together with the Friedmann equation $3\Mpl^2H^2=\rho_\phi+\rho_{\rm R}$ with $H$ being the Hubble parameter.
We then define the equality time $\widetilde t_{\rm RH}$ and temperature $\widetilde T_{\rm RH}$ by
\begin{equation}
\rho_{\rm R}(\widetilde t_{\rm RH})=\rho_\phi(\widetilde t_{\rm RH}),\qquad
\widetilde T_{\rm RH}=\left[\frac{30\rho_{\rm R}(\widetilde t_{\rm RH})}{\pi^2g_*(\widetilde t_{\rm RH})}\right]^{1/4}.
\label{eq:reheating-equality-temperature}
\end{equation}
We find $\widetilde T_{\rm RH}\simeq0.64\,T_{\rm RH}$. }

\begin{figure}
    \centering
    \includegraphics[width=0.49\linewidth]{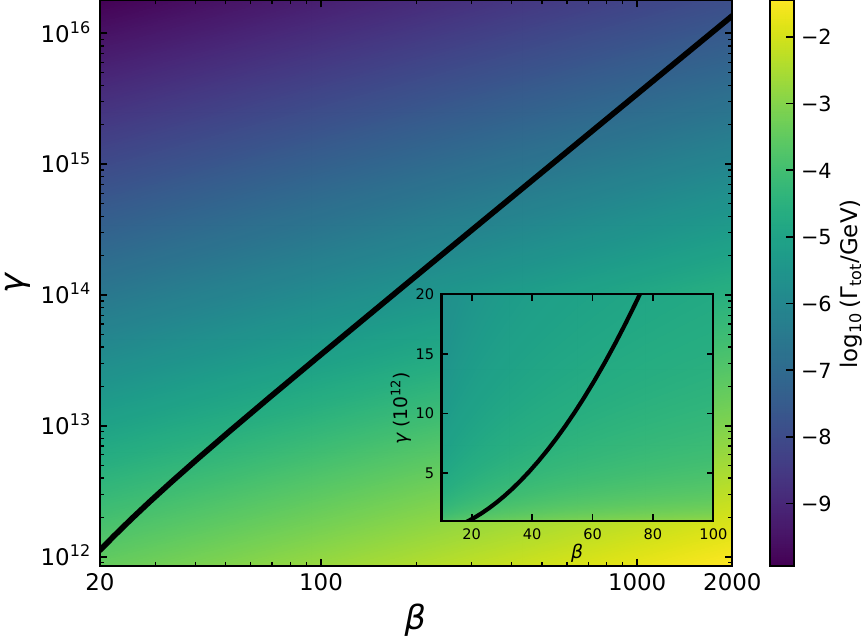}
\includegraphics[width=0.49\linewidth]{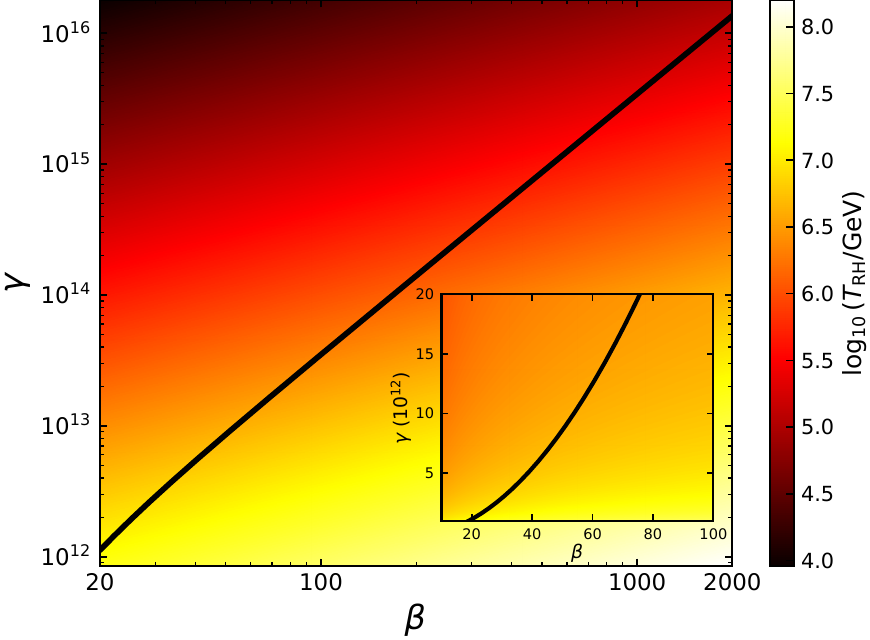}
    \caption{\textit{Left}: The density plot of total decay width of torsion on the $(\beta, \gamma)$ parameter space. \textit{Right}: The density plot of the reheating temperature $T_{\rm RH}$ on the $(\beta,\gamma)$ parameter space. In both plots, the black trajectory represents parameters that reproduce the observed density perturbation $A_s=2.1\times 10^{-9}$. {The trajectory uses continuous reheating and self-consistent pivot matching; the temperature background uses Eq.~\eqref{eq:trh-total-width}.}}
    \label{fig:placeholder}
\end{figure}

Fig.~\ref{fig:placeholder} shows total decay width and the reheating temperature on the $(\beta,\gamma)$ plane. The black trajectory represents parameters that reproduce the observed density perturbation, see next section for details.
{The background colors in the right panel show $T_{\rm RH}$, while the black trajectory is obtained from the self-consistent continuous-reheating calculation.}
In the next section we will see how possible observables depend on $T_{\rm RH}$ in order to distinguish the torsion-driven inflation models from others.

Before closing this section, we comment on the reheating in Starobinsky inflation~\cite{Vilenkin:1985md,Gorbunov:2010bn,Li:2021fao}.
In the Starobinsky model the scalaron couples to all non-conformal particles and the dominant decay channel is the 2 body decay into the Higgs boson pair with unsuppressed decay rate, $\Gamma \simeq m_\phi^3/(192\pi M_{\rm Pl}^2)$, unless the Higgs is conformally coupled to gravity.
Thus the reheating temperature is uniquely predicted to be $T_{\rm RH}\sim 5\times 10^9$\,GeV.
This is much higher than the prediction of the inflation models driven by torsion, as we will see below, and hence the reheating dynamics clearly distinguishes these models although the inflaton potential itself might look quite similar.

\section{Observational consequences} 
\label{sec:obs}

\subsection{Cosmic microwave background observables}

The reheating temperature logarithmically affects the e-folding number of inflation $(N_*)$ when the observable scales exit the horizon, which then affects the scalar spectral index of the curvature perturbation $(n_s)$ and the tensor-to-scalar ratio $(r)$.
We follow the standard procedure to calculate $n_s$ and $r$, and see the effect of the reheating temperature on them (see e.g. Refs.~\cite{Liddle:2000cg,Baumann:2009ds}).

The potential slow-roll parameters are defined by
\begin{equation}
\epsilon(\phi)\equiv
\frac{\Mpl^2}{2}\left(\frac{V'(\phi)}{V(\phi)}\right)^2,
\qquad
\eta(\phi)\equiv\Mpl^2\frac{V''(\phi)}{V(\phi)},
\label{eq:app-slow-roll-definitions}
\end{equation}
where prime denotes the derivative with respect to $\phi$.
By using the scalar potential (\ref{eq:app-exact-canonical-potential}), they are given by
\begin{align}
\epsilon(x)
&=\frac{4}{3}\frac{\cosh^2x}{(4\beta+\sinh x)^2},
\label{eq:app-exact-epsilon}\\
\eta(x)
&=\frac{4}{3}\left[
\frac{\cosh^2x}{(4\beta+\sinh x)^2}
+\frac{\sinh x}{4\beta+\sinh x}
\right],
\label{eq:app-exact-eta}
\end{align}
where $x$ is defined in Eq.~(\ref{eq:app-exact-x-definition}).
The two solutions of $\epsilon=1$ can be written as 
\begin{equation}
x_\pm=
\log\left(\frac{8\beta\pm\sqrt{64\beta^2-4/3}}
{2(2/\sqrt{3}-1)}\right).
\label{eq:app-slow-roll-boundaries}
\end{equation}
The lower solution $x_{\rm end}=x_-$ indicates the end of inflation on the plateau branch. The upper solution bounds the steep side and implies a finite maximum slow-roll duration. 
The e-folding number is evaluated by
\begin{equation}
N(\phi)\equiv\int_{t(\phi)}^{t_{\rm end}}H\,dt
\simeq\left[
3\beta\arctan(\sinh x)+\frac{3}{4}\log(\cosh x)
\right]_{x_{\rm end}}^{x}.
\label{eq:app-efold-definition}
\end{equation}
{The pivot matching follows from $k_*=a_*H_*$ and entropy conservation after reheating, $a(\widetilde t_{\rm RH})/a_0=(g_{*s,0}/g_{*s}(\widetilde t_{\rm RH}))^{1/3}\allowbreak T_0/\widetilde T_{\rm RH}$. Since the inflaton behaves as nonrelativistic matter after inflation ends, we have
\begin{equation}
\begin{split}
N_*\simeq{}&51.0+\frac13\ln\left(\frac{\widetilde T_{\rm RH}}{10^{10}\,{\rm GeV}}\right)
+\frac13\ln\left(\frac{H_{\rm end}}{10^{13}\,{\rm GeV}}\right)+\ln\left(\frac{H_*}{H_{\rm end}}\right),
\end{split}
\label{eq:N*}
\end{equation}}
{where $H_{\rm end}$ is the Hubble scale at the end of inflation, and we use $k_*=0.05\,{\rm Mpc}^{-1}$ and $g_{*}(\widetilde t_{\rm RH})=g_{*s}(\widetilde t_{\rm RH})=106.75$. Setting $H_*\simeq H_{\rm end}$ assumes a matter-like oscillation era followed by a sudden transition to radiation domination. The benchmarks use continuous reheating based on numerical calculation, including the end-of-inflation kinetic energy and the smooth transition to radiation domination, in the pivot matching.}
This determines the field value $\phi=\phi_*$ (or $x=x_*$) at $N=N_*$. Then we can calculate the power spectrum of the curvature perturbation $(A_s)$, scalar spectral index and the tensor-to-scalar ratio as follows:
\begin{equation}
A_s \simeq\frac{V_*}{24\pi^2\Mpl^4\epsilon_*},
\qquad
n_s=1-6\epsilon_*+2\eta_*,
\qquad
r=16\epsilon_*,
\label{eq:app-leading-slow-roll-observables}
\end{equation}
where $V_*=V(\phi_*)$, $\epsilon_*=\epsilon(\phi_*)$ and $\eta_*=\eta(\phi_*)$.
The Planck result indicates $n_s=0.965\pm0.004$~\cite{Planck:2018vyg},
while the recent Atacama Cosmology Telescope result combined with other observations shows preference for a larger value, $n_s=0.974\pm 0.003$~\cite{AtacamaCosmologyTelescope:2025blo}.

We have two model parameters $\beta$ and $\gamma$, which are equally expressed in terms of $q=\beta^2/\gamma$ and $\beta$.
The normalization of the curvature perturbation, $A_s \simeq 2.1\times 10^{-9}$~\cite{Planck:2018vyg}, fixes the value of $q$ $(\sim 10^{-10})$.
Then we are left with one parameter $\beta$. 
Since the decay rate (\ref{eq:total-width-factorization}) depends quadratically on $\beta$ (since $f_{\rm eff}$ (\ref{eq:feff-definition}) is roughly proportional to $\beta$ while the inflaton mass $m_\phi$ (\ref{eq:inflaton-mass}) is not very sensitive to $\beta$ once $q$ is fixed), one can predict $T_{\rm RH}$ for each chosen value of $\beta$.
Then one can predict $N_*$ and hence $n_s$ and $r$.\footnote{
    More precisely, we need to iterate the procedure until the parameters converge, since one already requires the value of $N_*$ in order to estimate $A_s$ and fix the value of $q$. 
}
Derived parameters for some benchmark points are summarized in Table~\ref{table:parameters}.

\begin{table}
\begin{center}
\begin{tabular}{|c|c|c|c|c|c|}
    \hline
    $\beta$ & $m_\phi$ & $\widetilde T_{\rm RH}$ &  $N_*$ & $n_s$ & $r$  \\ \hline
    $20$ & $3.76\times 10^{13}\,{\rm GeV}$ & $8.3\times10^6\,{\rm GeV}$ & $49.3$ & $0.971$ & $5.63\times 10^{-3}$ \\
    $30$ & $3.50\times 10^{13}\,{\rm GeV}$ & $5.0\times10^6\,{\rm GeV}$ & $49.1$ & $0.965$ & $4.88\times 10^{-3}$ \\
    $10^3$ & $3.39\times10^{13}\,{\rm GeV}$ & $1.4\times10^5\,{\rm GeV}$ & $47.9$ & $0.960$ & $4.55\times 10^{-3}$ \\ \hline
\end{tabular}
\caption{Parameters at several benchmark points.}
\label{table:parameters}
\end{center}
\end{table}

{Theoretically there is no upper bound for parameter $\beta$. For large-$\beta$, $f_{\rm eff}\propto\beta$ gives $\Gamma_{\rm tot}\propto\beta^{-2}$, so increasing $\beta$ lowers the reheating temperature and prolongs reheating, reducing $N_*$ and $n_s$. A lower limit on $n_s$ in this case translates into a lower limit on the reheating temperature. For $n_s\gtrsim0.960$, the self-consistent solutions give $\beta\lesssim10^3$ and $\widetilde T_{\rm RH}\gtrsim1.4\times10^5\,{\rm GeV}$. The large-$\beta$ benchmark with $n_s\simeq0.960$ lies within the quoted Planck interval.}

\begin{figure}[t]
\centering
\includegraphics[width=0.48\linewidth]{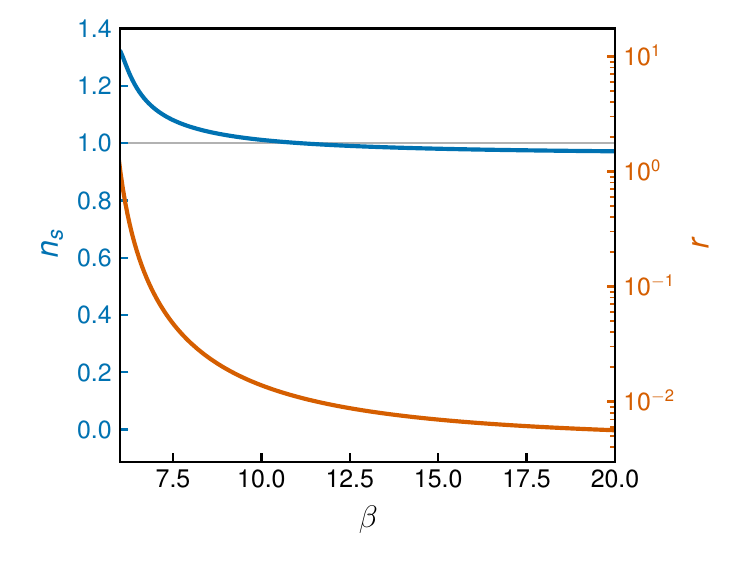}
\includegraphics[width=0.48\linewidth]{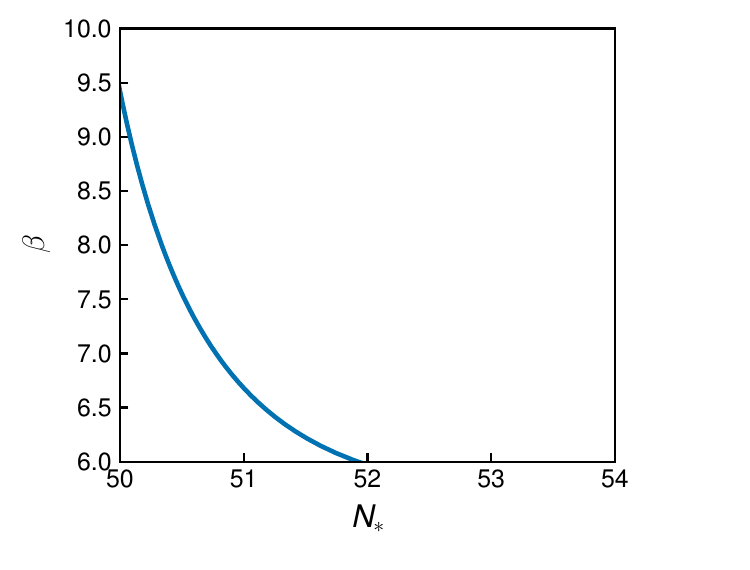}
\caption{{Self-consistent small-$\beta$ trajectory with $A_s=2.1\times10^{-9}$. Left: $n_s$ (blue, left axis) and $r$ (orange, right axis). Right: $\beta$ versus $N_*$. The curves use leading potential slow-roll expressions, slow-roll corrections become large toward $N_*\simeq51$.}}
\label{fig:small-beta-matching}
\end{figure}

{At smaller $\beta$, the plateau narrows and the pivot moves toward the steep side of the potential. In the limit $\beta\to0$ at fixed~$\gamma$,
\begin{equation}
V(\phi)\to\frac{\Mpl^4}{32\gamma}
\sinh^2\!\left(\sqrt{\frac23}\frac{\phi}{\Mpl}\right),\qquad
\epsilon(\phi)=\frac43\coth^2\!\left(\sqrt{\frac23}\frac{\phi}{\Mpl}\right)\geq\frac43,
\label{eq:beta-zero-no-slow-roll}
\end{equation}
so $\epsilon>1$ and slow-roll inflation is absent. Figure~\ref{fig:small-beta-matching} shows that $N_*\gtrsim50$ is reached only for $\beta\lesssim10$, where $n_s>1$; the solutions compatible with the observed tilt instead have $N_*<50$. The continuation toward $N_*\simeq51$ enters the regime of large slow-roll corrections. Reaching $N_*=50$--$60$ with an acceptable tilt requires a modified reheating history.}

\subsection{Gravitational waves}

\begin{figure}[htp!]
\centering
\includegraphics[width=0.7\textwidth]{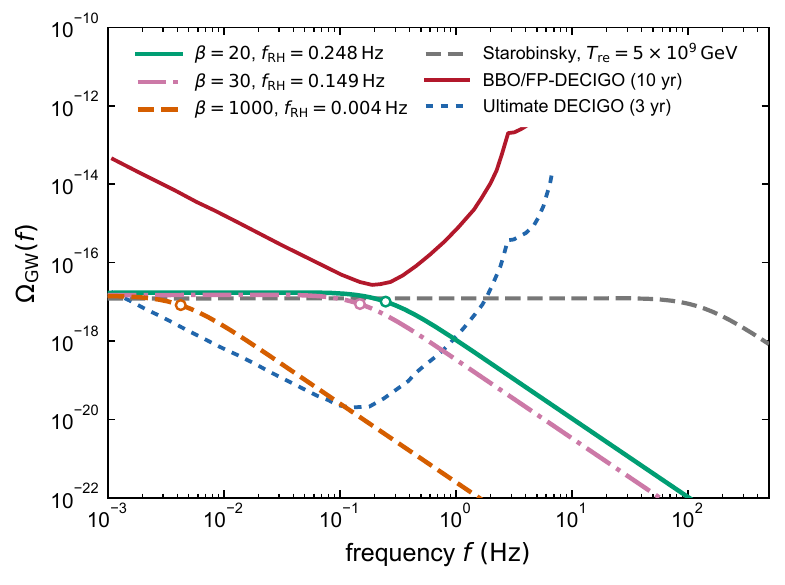}
\caption{Primordial gravitational wave spectrum for benchmark points $\beta=20,30,1000$. Sensitivity curves for DECIGO and ultimate-DECIGO are also shown.}
\label{fig:GWspectrum}
\end{figure}

Future observation of primordial gravitational waves with space laser interferometers, such as DECIGO~\cite{Seto:2001qf,Kawamura:2020pcg}, may directly determine the reheating temperature~\cite{Nakayama:2008ip,Nakayama:2008wy,Kuroyanagi:2008ye,Kuroyanagi:2011fy,Jinno:2014qka,Kuroyanagi:2014qza}.
The gravitational wave spectrum in terms of its density parameter $\Omega_{\rm GW}(f)$ is given by
\begin{equation}
\Omega_{\rm GW}(f)
=\frac{\Omega_{r,0}}{24}\,rA_s
\exp\left[n_t\ln\frac{f}{f_*}
+\frac12\alpha_t\ln^2\frac{f}{f_*}\right]
\left(\frac{g_*}{g_{*,0}}\right)
\left(\frac{g_{*s,0}}{g_{*s}}\right)^{4/3}\,\mathcal T_{\rm RH}^2\left(\frac{f}{f_{\rm RH}}\right),
\label{eq:gw-spectrum-high-frequency}
\end{equation}
with the slow-roll relations $n_t=-r/8$ and $\alpha_t=(r/8)(n_s-1+r/8)$ and we use $\Omega_{r,0}h^2=4.18\times10^{-5}$ for the radiation density parameter. The transfer function $\mathcal T_{\rm RH}(x)$ reflects the change of the slope of the spectrum due to the change of equation of state of the universe across the completion of reheating and it is approximately given by~\cite{Nakayama:2008wy},
\begin{equation}
\mathcal T_{\rm RH}^2(x)\simeq
\begin{cases}
1, & x\ll1,\\
x^{-2}, & x\gg1.
\end{cases}
\label{eq:gw-transfer-asymptotics}
\end{equation}
Modes with $f<f_{\rm RH}$ reenter during radiation domination and retain the
nearly scale-invariant inflationary spectrum. Modes with $f>f_{\rm RH}$ reenter during the matter-like inflaton oscillation era, and their fractional energy density acquires the $f^{-2}$ suppression.
Numerically,
\begin{equation}
f_{\rm RH}\simeq0.26\,{\rm Hz}
\left(\frac{T_{\rm RH}}{10^7\,{\rm GeV}}\right)
\left(\frac{g_*}{106.75}\right)^{1/6}.
\label{eq:reheating-knee-frequency}
\end{equation}
Fig.~\ref{fig:GWspectrum} shows the prediction of gravitational wave spectrum for our several benchmark points. Together shown are sensitivity curves of DECIGO or ultimate-DECIGO, taken from Ref.~\cite{Kuroyanagi:2011fy}.

\section{Conclusions and discussion}
\label{sec:conc}

Recently, there are growing interests on inflation driven by torsion degrees of freedom in extended gravity theories, partly due to the observational preference for higher scalar spectral index~\cite{He:2025bli,Katsoulas:2025srh,DiBenedetto:2026onj,DelGrosso:2026zbg}.
But the reheating of the inflation by dynamical torsion has not been explored in detail so far.
The derivative coupling nature of the torsion makes the analysis of decay processes a bit complicated.
While tree-level decays to fermion pair vanishes in the massless fermion limit, the 3 body decay picking up the top Yukawa coupling and the anomaly-induced 2 body decay into gauge bosons are substantial.
In the minimal reheating scenario, adopting $n_s\gtrsim0.960$ gives the benchmark $T_{\rm RH}\gtrsim10^5\,{\rm GeV}$. Lower reheating temperatures are also compatible with current cosmic microwave background observations for smaller $n_s$.

It is notable that it is lower than the prediction of the Starobinsky inflation model, $T_{\rm RH}\sim 5\times 10^9\,{\rm GeV}$. 
Thus these models can be observationally distinguished through the precision measurement of the scalar spectral index or the direct detection of primordial gravitational waves.

Let us comment on the leptogenesis scenario~\cite{Fukugita:1986hr}. 
Since the reheating temperature is relatively low, thermal leptogenesis does not work unless right-handed neutrino masses are degenerate.
However, as studied in Ref.~\cite{DiBenedetto:2026onj}, the inflaton can decay into right-handed neutrinos and nonthermal leptogenesis scenario may work~\cite{Asaka:1999yd,Asaka:1999jb}.
Comparing the partial decay rate to the fermion pair (\ref{Gamma_phi_to_ff}) with (\ref{eq:total-width-factorization}), the decay into right-handed neutrinos can be dominant if $m_N \gtrsim 0.04\times m_\phi$. 
Otherwise, the branching ratio for the right-handed neutrino production is suppressed and the efficiency of the leptogenesis is reduced.
In order to evaluate the lepton or baryon asymmetry of the universe, the evaluation of the branching ratio is essential and our study gives a robust foundation for it.

Although we mainly focused on the case of torsion as an inflaton and studied its decay and reheating dynamics, our calculations can be generally applied to derivatively-coupled scalar particles.
One may identify such a derivatively-coupled scalar as an axion-like particle that constitutes the dark matter of the universe.
In this case it is often much lighter than all the fermions, and the decay width calculation is different from the inflaton case, but we also provided a useful formula applicable even in such a case: see Appendix~\ref{sec:loop2body}.

Finally we comment on the high frequency gravitational waves possibly produced by the inflaton dynamics during the reheating.
It is known that the perturbative inflaton decay is always accompanied with the bremsstrahlung emission of the graviton~\cite{Nakayama:2018ptw,Barman:2023ymn,Barman:2023rpg,Hu:2024awd,Hu:2024bha}. 
In our model, we have checked that the 3 body decay into the fermion pair plus the graviton vanishes.
Thus the dominant bremsstrahlung contribution may come from the graviton emission associated with the anomaly-induced decay or the Yukawa-induced 3 body decay. 
The graviton pair production from inflaton annihilation~\cite{Ema:2015dka,Ema:2016hlw,Ema:2020ggo,Choi:2024ilx,Bernal:2025lxp,Xu:2025wjq,Mudrunka:2026kgm,Wang:2026ule,Wang:2026pff} should be the same as conventional inflation models.
Some complexities may arise for evaluating possible inflaton decay into the graviton pair~\cite{Ema:2021fdz,Mudrunka:2023wxy,Tokareva:2023mrt,Strumia:2025dfn,Nakayama:2025xkn}, or contributions from scattering of high-energy particles with the inflaton~\cite{Xu:2024fjl,Bernal:2025lxp,Xu:2025wjq}.
Note also that scattering of Standard Model particles in thermal bath produces stochastic gravitational waves at high frequency range~\cite{Ghiglieri:2015nfa,Ghiglieri:2020mhm,Ringwald:2020ist,Ghiglieri:2022rfp}, although relatively low reheating temperature in our model should suppress its total amount.
We leave detailed calculations of high frequency gravitational waves for a future work.

\section*{Acknowledgment}

This work was supported by World Premier International Research Center Initiative (WPI), MEXT, Japan.
This work was also supported by JSPS KAKENHI (Grant Number 24K07010 [KN], 26K00695 [KN], 26H00403 [KN]).

\appendix
\section{Detailed calculation of torsion decay}
\label{app:decay}

The derivative interaction in Eq.~\eqref{eq:canonical-torsion-coupling}
has an equivalent description in terms of nonderivative Yukawa operators
and anomalous gauge operators.  This form makes the nonvanishing decay
channels manifest.  We work above the electroweak scale and neglect all
Standard Model masses in the kinematics.
For the Yukawa-induced channels, we keep only the top-quark contribution
because $y_t$ is the largest Standard Model Yukawa coupling, contributions
from lighter fermions are suppressed by $|y_f/y_t|^2$ in the decay widths.

\subsection{Torsion couplings}
\label{app:torsion_coupling}

First we collect the torsion couplings to the Standard Model chiral multiplets into
the compact current form,
\begin{align}
&\Lag_{\phi J}
={}\frac{\partial_\mu\phi}{f_{\rm eff}}J_T^\mu,\\
&J_T^\mu
={}\sum_{i=1}^{3}\left(
 \bar u_R^i\gamma^\mu u_R^i
+\bar d_R^i\gamma^\mu d_R^i
+\bar e_R^i\gamma^\mu e_R^i
-\bar Q_L^i\gamma^\mu Q_L^i
-\bar L_L^i\gamma^\mu L_L^i
\right).
\label{eq:app-current-interaction}
\end{align}
The relative signs follow from the axial torsion coupling.  We don't assume the presence of right-handed neutrino here, but it is straightforward to include it.  Below we use the shorthand $f=f_{\rm eff}$, and define the Standard Model Yukawa operator
\begin{equation}
{\cal O}_Y
=\bar Q_L\widetilde H Y_u u_R
+\bar Q_L H Y_d d_R
+\bar L_L H Y_e e_R,
\qquad
\Lag_Y=-{\cal O}_Y-{\cal O}_Y^\dagger .
\label{eq:app-yukawa-operator}
\end{equation}
The most direct derivation of the current divergence follows from the chiral
transformation generated by $J_T^\mu$.  For a transformation parameter
$\alpha$, the fields transform as
\begin{equation}
u_R,d_R,e_R\longrightarrow e^{+i\alpha}(u_R,d_R,e_R),
\qquad
Q_L,L_L\longrightarrow e^{-i\alpha}(Q_L,L_L),
\qquad
H\longrightarrow H.
\label{eq:app-torsion-chiral-transformation}
\end{equation}
Equivalently, the generation-space charge matrices used below are
$c_u=c_d=c_e=\boldsymbol{1}$ and
$c_Q=c_L=-\boldsymbol{1}$.
In the top sector this reduces to $t_R\to e^{i\alpha}t_R$ and
$Q_L\to e^{-i\alpha}Q_L$, and we have the non-conservation of the currents at classical level,
\begin{equation}
\left.\partial_\mu J_T^\mu\right|_Y
=2i\left({\cal O}_Y-{\cal O}_Y^\dagger\right).
\label{eq:app-yukawa-current-divergence}
\end{equation}

For comparison, Eq.~\eqref{eq:app-yukawa-current-divergence} can be checked
explicitly with the fermion equations of motion.   The relevant classical equations of motion for fermions 
are
\begin{align}
i\slashed D Q_L
&=\widetilde H Y_u u_R+H Y_d d_R,
&
i\slashed D L_L
&=H Y_e e_R,
\notag\\
i\slashed D u_R
&=Y_u^\dagger\widetilde H^\dagger Q_L,
&
i\slashed D d_R
&=Y_d^\dagger H^\dagger Q_L,
\notag\\
i\slashed D e_R
&=Y_e^\dagger H^\dagger L_L,
\label{eq:app-sm-fermion-eom}
\end{align}
together with their conjugates.   Since we have relation 
\begin{equation}
\partial_\mu\!\left(\bar\psi\gamma^\mu\psi\right)
=\left(D_\mu\bar\psi\right)\gamma^\mu\psi
+\bar\psi\gamma^\mu D_\mu\psi,
\label{eq:app-current-bilinear-divergence}
\end{equation}
substitution of Eq.~\eqref{eq:app-sm-fermion-eom} and its conjugate into
Eq.~\eqref{eq:app-current-bilinear-divergence} gives two equal contributions, and their sum reproduces
Eq.~\eqref{eq:app-yukawa-current-divergence}.

The classical equations above of motion do not account for the gauge anomaly coming from the chiral
anomalous Ward identity, or equivalently the Jacobian of the corresponding
local chiral field redefinition, yields
\begin{align}
\left.\partial_\mu J_T^\mu\right|_{\rm anom}
=-\left[
\frac{\alpha_s}{4\pi}C_3G_{\mu\nu}^A\widetilde G^{A\mu\nu}
+\frac{\alpha_2}{4\pi}C_2W_{\mu\nu}^I\widetilde W^{I\mu\nu}
+\frac{\alpha_1}{4\pi}C_1B_{\mu\nu}\widetilde B^{\mu\nu}
\right],
\label{eq:app-anomalous-current-divergence}
\end{align}
where $\alpha_s = g_s^2/(4\pi)$, $\alpha_2=g_2^2/(4\pi)$, $\alpha_1=g_1^2/(4\pi)$ with $g_s, g_2$ and $g_1$ being the SU(3), SU(2)$_L$ and U(1)$_Y$ gauge coupling constants. 
The numerical factor $C_1, C_2$ and $C_3$ will be derived later.

Finally, integration by parts gives
$\Lag_{\phi J}=-\phi\,\partial_\mu J_T^\mu/f$ up to a total derivative.
Combining Eqs.~\eqref{eq:app-yukawa-current-divergence} and
\eqref{eq:app-anomalous-current-divergence} yields
\begin{align}
\Lag_{\phi J}
={}&-\frac{2i\phi}{f}
\left({\cal O}_Y-{\cal O}_Y^\dagger\right)
\notag\\
&+\frac{\phi}{f}\left[
\frac{\alpha_s}{4\pi}C_3G_{\mu\nu}^A\widetilde G^{A\mu\nu}
+\frac{\alpha_2}{4\pi}C_2W_{\mu\nu}^I\widetilde W^{I\mu\nu}
+\frac{\alpha_1}{4\pi}C_1B_{\mu\nu}\widetilde B^{\mu\nu}
\right],
\label{eq:app-equivalent-decay-basis}
\end{align}
with $\widetilde X^{\mu\nu}=\epsilon^{\mu\nu\rho\sigma}
X_{\rho\sigma}/2$. 

The first diagram in Fig.~\ref{fig:app-equivalent-vertices} is the local interaction generated by the Yukawa part of the current divergence, leading to the characteristic three body decay channel.  The second is the anomalous gauge interaction. Its triangle representation and finite-mass threshold correction are evaluated in
Sec.~\ref{sec:loop2body}.

\subsection{2 body decay to gauge bosons}
\label{app:2body_to_gauge}

First let us derive the decay rate induced by the second line of (\ref{eq:app-equivalent-decay-basis}), i.e., the anomaly term. To do so, we need to know values of $C_1, C_2$ and $C_3$.
We normalize the anomaly coefficient of one Dirac color triplet to unity.
The non-Abelian coefficients are then
\begin{align}
C_3
&=T_F\operatorname{Tr}_g\!\left(c_u+c_d-N_Lc_Q\right)
=\frac{1}{2}\times3\,[1+1-2(-1)]=6,
\notag\\
C_2
&=-T_F\operatorname{Tr}_g\!\left(N_cc_Q+c_L\right)
=-\frac{1}{2}\times3\,[3(-1)+(-1)]=6,
\label{eq:app-nonabelian-anomaly-coefficients}
\end{align}
where $T_F=1/2$ comes from Dynkin index of the non-Abelian group, $\operatorname{Tr}[T^aT^b]=T_F\delta^{ab}=1/2\delta^{ab}$, $N_c=3$, $N_L=2$, and $\operatorname{Tr}_g$ is the
generation trace. With
$Y_Q=1/6$, $Y_u=2/3$, $Y_d=-1/3$, $Y_L=-1/2$, and $Y_e=-1$, this gives
\begin{align}
C_1
&=\operatorname{Tr}_g\!\left[
N_c\left(Y_u^2c_u+Y_d^2c_d-N_LY_Q^2c_Q\right)
+Y_e^2c_e-N_LY_L^2c_L
\right]
\notag\\
&=3\times\left[
3\times\bigg(\left(\frac{2}{3}\right)^2
+\left(-\frac{1}{3}\right)^2+
2\left(\frac{1}{6}\right)^2\bigg)
+(-1)^2+2\left(-\frac{1}{2}\right)^2
\right]=10.
\label{eq:app-hypercharge-anomaly-coefficient}
\end{align}
The group multiplicities enter only in the final-state sum,
\begin{equation}
d_3C_3^2=8\times6^2=288,
\qquad
d_2C_2^2=3\times6^2=108,
\qquad
d_1C_1^2=10^2=100,
\label{eq:app-anomaly-group-sums}
\end{equation}
with $d_3=8$ being the number of gluons, $d_2=3$ the number of weak isospin
projections, and $d_1=1$ the number of hypercharges.  For a gauge group $a$, Eq.~\eqref{eq:app-equivalent-decay-basis} gives
\begin{equation}
\Gamma(\phi\to V_aV_a)
=\frac{d_aC_a^2\alpha_a^2}{64\pi^3}
\frac{m_\phi^3}{f^2}.
\label{eq:app-individual-anomaly-width}
\end{equation}
Summing over the three Standard Model gauge groups, we obtain
\begin{equation}
\Gamma_{\rm anom}
=\frac{m_\phi^3}{64\pi^3f^2}
\left(288\alpha_s^2+108\alpha_2^2+100\alpha_1^2\right).
\label{eq:app-total-anomaly-width}
\end{equation}

\subsection{3 body decay to fermion pair plus Higgs boson}

The first line of (\ref{eq:app-equivalent-decay-basis}) is responsible for the 3 body decay.
The top-Yukawa part of Eq.~\eqref{eq:app-equivalent-decay-basis} is
\begin{equation}
\Lag_{\phi,t}^{(4)}
=\frac{2iy_t^*}{f}\,\phi\,
\bar t_R\widetilde H^\dagger Q_L+\mathrm{h.c.}
\label{eq:app-contact-interaction}
\end{equation}
Here $\widetilde H^\dagger Q_L=H_2t_L-H_1b_L$.  The displayed operator is
the $\phi H_i^\dagger\bar t_RQ_L$ contact vertex used in
Fig.~\ref{fig:app-hhv-current-diagrams}.
Consider one component,
$\phi(p)\to H_a^\dagger(k_H)+Q_a(k_Q)+\bar t_R(k_t)$.  The contact diagram in
Fig.~\ref{fig:app-equivalent-vertices}(a) gives
\begin{equation}
{\cal M}_a
=2i\frac{y_t^*}{f}\bar u_Q(k_Q)P_Rv_t(k_t),
\qquad
\sum_{\rm spins}|{\cal M}_a|^2
=4\frac{|y_t|^2}{f^2}s_{Qt},
\label{eq:app-contact-amplitude}
\end{equation}
where $s_{Qt}=(k_Q+k_t)^2$.  For three massless final states,
\begin{equation}
d\Gamma_a
=\frac{N_c}{256\pi^3m_\phi^3}
\sum_{\rm spins}|{\cal M}_a|^2,ds_{Qt}\,ds_{Ht},
\qquad
\int ds_{Qt}\,ds_{Ht}\,s_{Qt}=\frac{m_\phi^6}{6}.
\label{eq:app-contact-phase-space}
\end{equation}
The width of one weak component is therefore
\begin{equation}
\Gamma_a
=\frac{N_c|y_t|^2}{384\pi^3}\frac{m_\phi^3}{f^2}.
\end{equation}
The doublet contraction in Eq.~\eqref{eq:app-contact-interaction} contains two
weak components.  Its Hermitian conjugate supplies two charge-conjugate
channels, so their sum is
\begin{equation}
\Gamma_{\rm contact}
=4\Gamma_a
=\frac{N_c|y_t|^2}{96\pi^3}\frac{m_\phi^3}{f^2}
=\frac{|y_t|^2}{32\pi^3}\frac{m_\phi^3}{f^2},
\qquad N_c=3.
\label{eq:app-contact-width}
\end{equation}

\subsection{2-body decay to gauge bosons: the case of massive fermion}
\label{sec:loop2body}

Here let us keep the fermion mass finite and estimate the fermion mass dependence on the decay rate into gauge boson pair. The underlying derivative interaction has the same operator structure as the axion-fermion coupling, whose loop-induced decay into gauge bosons was studied in Refs.~\cite{Bauer:2017ris,Buen-Abad:2021fwq,Liu:2022tqn}. For our application to the inflaton we can safely assume massless fermions, but keeping the fermion mass will be helpful for understanding how the anomaly contribution and one-loop contribution are related (see also Ref.~\cite{He:2025fij}). 
Also, if we introduce heavy fermions or want to consider very light distortion for some other phenomenological applications, we need to clearly keep the fermion mass finite.

\begin{figure}[htp!]
\centering
\begin{minipage}[c]{0.46\textwidth}
\centering
\begin{tikzpicture}[baseline=(current bounding box.center)]
\begin{feynman}
\vertex (phi) at (-1.55,0) {$\phi$};
\vertex[dot] (vphi) at (0,0) {};
\vertex[dot] (vone) at (1.40,0.90) {};
\vertex[dot] (vtwo) at (1.40,-0.90) {};
\vertex (gone) at (3.10,0.90) {};
\vertex (gtwo) at (3.10,-0.90) {};
\diagram*{
  (phi) -- [scalar, momentum=\(p\)] (vphi),
  (vphi) -- [momentum=\(p+k\)] (vone),
  (vone) -- [momentum=\(p+k-p_1\)] (vtwo),
  (vtwo) -- [momentum=\(k\)] (vphi),
  (vone) -- [photon, momentum=\(p_1\)] (gone),
  (vtwo) -- [photon, momentum'=\(p_2\)] (gtwo)
};
\end{feynman}
\end{tikzpicture}
\par\vspace{-0.4em}\textbf{(a)}
\end{minipage}\hfill
\begin{minipage}[c]{0.46\textwidth}
\centering
\begin{tikzpicture}[baseline=(current bounding box.center)]
\begin{feynman}
\vertex (phi) at (-1.55,0) {$\phi$};
\vertex[dot] (vphi) at (0,0) {};
\vertex[dot] (vone) at (1.40,0.90) {};
\vertex[dot] (vtwo) at (1.40,-0.90) {};
\vertex (gone) at (3.25,1.30) {};
\vertex (gtwo) at (3.25,-1.30) {};
\diagram*{
  (phi) -- [scalar, momentum=\(p\)] (vphi),
  (vphi) -- [momentum=\(p+k\)] (vone),
  (vone) -- (vtwo),
  (vtwo) -- [momentum=\(k\)] (vphi),
  (vone) -- [photon] (gtwo),
  (vtwo) -- [photon] (gone)
};
\end{feynman}
\draw[->] (2.70,0.42) -- (3.05,0.84);
\node at (3.18,0.52) {$p_1$};
\draw[->] (2.70,-0.42) -- (3.05,-0.84);
\node at (3.18,-0.52) {$p_2$};
\end{tikzpicture}
\par\vspace{-0.4em}\textbf{(b)}
\end{minipage}
\caption{Loop-induced decay into gauge boson pair.}
\label{fig:1loop}
\end{figure}
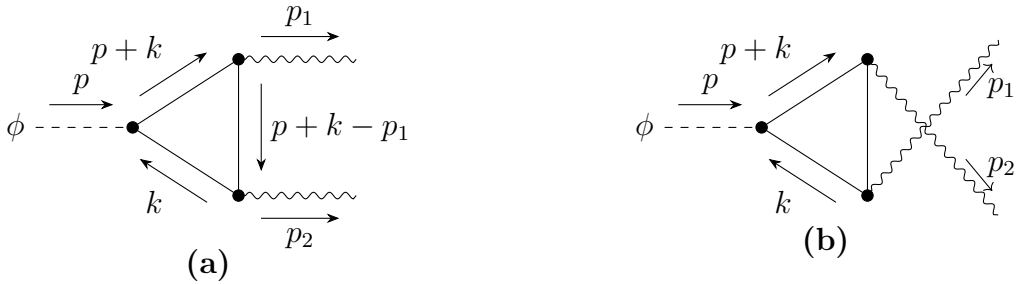

Again let us start from the Lagrangian of the form
\begin{align}
	\mathcal L = \frac{\partial_\mu \phi}{f}\sum_i\overline\psi_i\gamma_\mu\gamma_5\psi_i,
    \label{delphi_psi5psi}
\end{align}
and suppose that $\psi_i$ has a charge $q_i$ under U(1) gauge group. 
Chiral rotations of the fermion, $\psi_{L_i} \to e^{i\theta_L}\psi_{Li}$ and $\psi_{R_i} \to e^{i\theta_R}\psi_{Ri}$ with $\theta_L=-\theta_R=\phi/f$, can remove this term. Instead, following terms appear:
\begin{align}
	\mathcal L =-\frac{2i\phi}{f}\sum_i m_i \overline\psi_i\gamma_5\psi_i + \sum_i\frac{\alpha_e q_i^2}{2\pi} \frac{\phi}{f} F_{\mu\nu}\widetilde F^{\mu\nu}.
    \label{phi_psi5psi}
\end{align}
where $\alpha_e=e^2/(4\pi)$ with $e$ being the gauge coupling constant.
Let us consider the decay process $\phi\to VV$ in this basis, where $V$ denotes the (massless) U(1) gauge boson. There are tree-level and one-loop contributions.  The former comes from the anomaly term, and the latter from the fermion loop, shown in Fig.~\ref{fig:1loop}:
\begin{align}
	i\mathcal M = i \mathcal M_{\rm tree} +  i \mathcal M_{\rm loop}.
\end{align}

The one-loop contribution is calculated as (see Fig.~\ref{fig:1loop} for momentum assignments)
\begin{align}
	\mathcal M_{\rm loop} =& \sum_i\frac{2(eq_i)^2 m_i}{f} \int \frac{d^4k}{(2\pi)^4}{\rm Tr}\left[ \gamma_5\frac{\slashed{p}+\slashed{k}+m_i}{(p+k)^2-m_i^2}	\gamma^\mu \frac{\slashed{p_2}+\slashed{k}+m_i}{(p_2+k)^2-m_i^2}\gamma^\nu \frac{\slashed{k}+m_i}{k^2-m_i^2}\right. \nonumber\\
	& +\left.  \gamma_5\frac{\slashed{p}+\slashed{k}+m_i}{(p+k)^2-m_i^2}	\gamma^\mu \frac{\slashed{p_1}+\slashed{k}+m_i}{(p_1+k)^2-m_i^2}\gamma^\nu \frac{\slashed{k}+m_i}{k^2-m_i^2}\right]\epsilon^*_\mu(p_1)\epsilon^*_\nu(p_2).
    \label{iM_loop}
\end{align}
It is noticed that all the nonzero terms involve
\begin{align}
	{\rm Tr}\left[ \gamma_5 \gamma^\rho\gamma^\mu\gamma^\sigma\gamma^\nu\right] = -4i \epsilon^{\rho\mu\sigma\nu}.
\end{align}
After some calculations, we find
\begin{align}
	\mathcal M_{\rm loop} &= \sum_i \frac{2(eq_i)^2}{f} \int \frac{d^4k}{(2\pi)^4} \frac{-8im_i^2\epsilon^{\mu\nu\rho\sigma}\,p_{1\rho} p_{2\sigma} \epsilon^*_\mu(p_1)\epsilon^*_\nu(p_2)}{\left[(p_1+k)^2-m_i^2\right]\left[(p_2-k)^2-m_i^2\right](k^2-m_i^2)}\nonumber\\
	&=-\sum_i \frac{(eq_i)^2}{f}32im_i^2\epsilon^{\mu\nu\rho\sigma}\,p_{1\rho} p_{2\sigma} \epsilon^*_\mu(p_1)\epsilon^*_\nu(p_2) \int \frac{d^4\ell}{(2\pi)^4}\int dxdydz \frac{\delta(x+y+z-1)}{\left(\ell^2-\Delta_i\right)^3} \nonumber\\
	&=-\sum_i \frac{(eq_i)^2}{f}\frac{m_i^2}{\pi^2} \epsilon^{\mu\nu\rho\sigma}\,p_{1\rho} p_{2\sigma} \epsilon^*_\mu(p_1)\epsilon^*_\nu(p_2) \int_0^1 dx \int_0^{1-x} dy \frac{1}{\Delta_i},
\end{align}
where $\ell = k + xp_1-yp_2$, $\Delta_i = -2xy(p_1\cdot p_2)+m_i^2$, using $(p_1)^2=(p_2)^2=0$.
Adding the tree-level contribution, we find
\begin{align}
	\mathcal M = \sum_i\frac{(eq_i)^2}{2\pi^2 f}\epsilon^{\mu\nu\rho\sigma}\,p_{1\rho} p_{2\sigma} \epsilon^*_\mu(p_1)\epsilon^*_\nu(p_2) \left[1- \int_0^1 dx \int_0^{1-x} dy \frac{2m_i^2}{\Delta_i} \right].
	\label{amp_agg}
\end{align}
The integral is evaluated as
\begin{align}
	I_i(\eta) &\equiv \int_0^1 dx \int_0^{1-x} dy \frac{2m_i^2}{\Delta_i - i\epsilon} =  \int_0^1 dx \int_0^{1-x} dy \frac{2}{1-4xy/\eta_i- i\epsilon} \nonumber\\
	&=\begin{cases}
		\displaystyle \eta_i \, {\rm arcsin^2} \left(\frac{1}{\sqrt{\eta_i}}\right) & (\eta_i > 1) \\
		\displaystyle -\frac{\eta_i}{4}\left[ \ln\left(\frac{1+\sqrt{1-\eta_i}}{1-\sqrt{1-\eta_i}}\right)-i\pi \right]^2 & (\eta_i < 1)
	\end{cases},
    \label{I_eta}
\end{align}
where we have defined $\eta_i \equiv 4m_i^2/m_\phi^2$.

\paragraph{Heavy fermion limit}

First let us consider the case of $\eta_i \gg 1$ $(m_\phi^2\ll 4m_i^2)$, i.e., fermions are heavy enough to be integrated out. 
In this case we can easily evaluate the integral as
\begin{align}
	 I_i(\eta) = \int_0^1 dx \int_0^{1-x} dy \frac{2m_i^2}{\Delta_i}  \simeq 1 + \frac{p^2}{12 m_i^2}.
\end{align}
Thus the anomaly contribution is canceled out. The effective scalar-gauge-boson interaction is then written as~\cite{Nakayama:2014cza}
\begin{align}
	\mathcal L_{\rm eff} \simeq -\sum_{i{\rm:heavy}}\frac{\alpha_e q_i^2}{24\pi f} \frac{\partial^2 \phi}{m_i^2} F_{\mu\nu}\widetilde F^{\mu\nu},
\end{align}
where the summation is taken over the heavy fermion $\eta_i \gg 1$.
The decay rate from heavy fermion contribution is then given by
\begin{align}
	\Gamma(\phi\to VV) = \frac{1}{32\pi m_\phi}\sum_{\rm pol}\left|\mathcal M\right|^2 = \frac{\alpha_e^2}{2304\pi^3} \left(\sum_{i{\rm :heavy}}\frac{q_i^2}{m_i^2}\right)^2\frac{m_\phi^7}{f^2}.
\end{align}

\paragraph{Light fermion limit}

Next let us consider the opposite case: $\eta_i \ll 1$ $(m_\phi^2 \gg 4m_i^2)$. Then the integral is suppressed as $I_i(\eta) \sim \mathcal O(\eta_i) \ll 1$.
Thus the integral is negligible in the amplitude (\ref{amp_agg}), meaning that the anomaly term solely contributes to the total amplitude. The decay rate is
\begin{align}
	\Gamma(\phi\to VV) = \frac{\alpha_e^2}{16\pi^3} \left(\sum_{i{\rm :light}} q_i^2\right)^2 \frac{m_\phi^3}{f^2},
    \label{G_phi2g_anomaly}
\end{align}
where the summation is taken over light fermion species, $\eta_i < 1$.
This is independent of the fermion mass $m_i$. It should be remarked that the 2-body decay rate $\phi\to\psi\overline\psi$ vanishes in the limit $m_i\to 0$, but the one-loop decay rate remains constant.
Combining both cases, we obtain the following expression:
\begin{align}
	\Gamma(\phi\to VV) \simeq \frac{\alpha_e^2}{16\pi^3} \left(\sum_{i{\rm :light}} q_i^2
    -\sum_{i{\rm :heavy}}\frac{q_i^2 m_\phi^2}{12m_i^2}\right)^2 \frac{m_\phi^3}{f^2}.
\end{align}
If some of the fermions have masses comparable to the scalar ($m_i\sim m_\phi$), one should use the full integral expression (\ref{I_eta}).
All these results are consistent with e.g. Ref.~\cite{Bauer:2017ris}.

Notice the difference from the case of conventional axion models, which are usually defined by only the first term of (\ref{phi_psi5psi}) in the model basis. In such a case, after the chiral rotation to remove it, we find a derivative coupling (\ref{delphi_psi5psi}) plus the anomaly term, i.e., the second term of (\ref{phi_psi5psi}).
It is rather easy to calculate the decay rate $\phi\to VV$ in the original basis, and we find that the decay rate is given by (\ref{G_phi2g_anomaly}) for $m_\phi^2\ll 4m_i^2$, while the rate $\phi\to VV$ is suppressed by the ratio $(4m_i^2/m_\phi^2)^2$ for $m_\phi^2\gg 4m_i^2$.
Therefore, both $\phi\to\psi\overline\psi$ and $\phi\to VV$ rates vanish in the $m_i\to 0$ limit.

\paragraph{The case of non-Abelian gauge boson}

So far we have considered a $\phi$ decay to U(1) gauge boson pair. Let us suppose that some of $\psi_i$ in (\ref{delphi_psi5psi}) are fundamental representations of SU($N$) and others are singlet.
Then, after the chiral rotations we have
\begin{align}
	\mathcal L =-\frac{2i\phi}{f}\sum_i m_i \overline\psi_i\gamma_5\psi_i + \sum_i\frac{\alpha_e q_i^2}{2\pi} \frac{\phi}{f} F_{\mu\nu}\widetilde F^{\mu\nu}
    +\sum_i \frac{\alpha_g C_i}{2\pi} \frac{\phi}{f} G_{\mu\nu}^a\widetilde G^{\mu\nu a},
    \label{phi_psi5psi_NonAbelian}
\end{align}
where $\alpha_g=g^2/(4\pi)$ with $g$ being the SU($N$) gauge coupling constant and we assign $C_i=1/2$ for fundamental representations and $C_i=0$ for others.
The one-loop calculation of the amplitude for the process $\phi$ decay to the SU($N$) gauge boson pair $(\phi\to gg)$ is almost parallel: we should only multiply the factor ${\rm Tr}(T^aT^b) =\frac{1}{2}\delta^{ab}$ to the amplitude (\ref{iM_loop}) if the fermion in the loop is fundamental representation.
As a result, the decay rate is estimated as
\begin{align}
	\Gamma(\phi\to gg) \simeq \frac{(N^2-1)\alpha_g^2}{16\pi^3} \left(\sum_{i{\rm :light}} C_i -\sum_{i{\rm :heavy}}\frac{C_i m_\phi^2}{12m_i^2}\right)^2 \frac{m_\phi^3}{f^2}.
\end{align}
In the phenomenological applications discussed in the main text, $\phi$ is inflaton and all the Standard Model fermions are regarded as massless. Thus only the first term is relevant.
By correctly counting the Standard Model particle content, we obtain the same result as Eq.~(\ref{eq:app-total-anomaly-width}).

\subsection{3 body decay to fermion pair plus Higgs boson: original basis}

\begin{figure}[htp!]
\centering
\begin{minipage}[c]{0.46\textwidth}
\centering
\begin{tikzpicture}[baseline=(current bounding box.center)]
\begin{feynman}
\vertex (phi) at (-1.65,0) {$\phi$};
\vertex[dot] (current) at (0,0) {};
\vertex (q) at (1.75,0.85) {$t_L$};
\vertex[dot] (yukawa) at (1.20,-0.65) {};
\vertex (tb) at (2.65,-0.65) {$\bar t_R$};
\vertex (h) at (1.20,-1.65) {$H_2$};
\diagram*{
  (phi) -- [scalar] (current),
  (yukawa) -- [fermion, edge label=\(t_L\)] (current),
  (current) -- [fermion] (q),
  (tb) -- [fermion] (yukawa),
  (yukawa) -- [scalar] (h)
};
\end{feynman}
\end{tikzpicture}
\par\vspace{-0.4em}\textbf{(a)}
\end{minipage}\hfill
\begin{minipage}[c]{0.46\textwidth}
\centering
\begin{tikzpicture}[baseline=(current bounding box.center)]
\begin{feynman}
\vertex (phi) at (-1.65,0) {$\phi$};
\vertex[dot] (current) at (0,0) {};
\vertex (tb) at (1.75,-0.85) {$\bar t_R$};
\vertex[dot] (yukawa) at (1.20,0.65) {};
\vertex (q) at (2.65,0.65) {$t_L$};
\vertex (h) at (1.20,1.65) {$H_2$};
\diagram*{
  (phi) -- [scalar] (current),
  (tb) -- [fermion] (current),
  (current) -- [fermion, edge label'=\(t_R\)] (yukawa),
  (yukawa) -- [fermion] (q),
  (yukawa) -- [scalar] (h)
};
\end{feynman}
\end{tikzpicture}
\par\vspace{-0.4em}\textbf{(b)}
\end{minipage}
\caption{The diagrams for
$\phi\to H_2^\dagger t_L\bar t_R$ at original basis.  The $\phi$ vertex is inserted on the
$Q_L$ line in panel (a) and on the $t_R$ line in panel (b).  In each panel,
the vertex adjacent to $\phi$ is the derivative-current insertion and the
second vertex is the top-Yukawa interaction.  The charge-conjugate diagrams
are implicit.}
\label{fig:app-three-body-original-basis}
\end{figure}
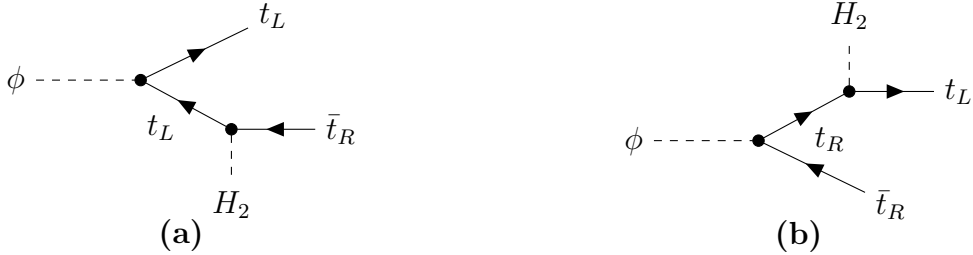

We now verify the Yukawa-assisted decay directly in the original
basis of Eq.~\eqref{eq:app-current-interaction}.  This provides a
basis-independence check of the contact-interaction calculation above. Consider the process 
$\phi(p)\to H(k_H)+Q(k_Q)+\bar t_R(k_t)$.
In the derivative-current basis, the tree-level amplitude is the sum of two
diagrams in which the $\phi$ vertex is attached to the $t_R$ or $Q_L$
fermion line adjacent to the top-Yukawa vertex, as shown in
Fig.~\ref{fig:app-three-body-original-basis}.

The top Yukawa interaction in the unbroken phase is
\begin{equation}
{\cal L}_{y_t}
=-y_t\bar Q_L\widetilde Ht_R+{\rm h.c.}
=-y_t\left(
\bar t_LH_2^*t_R-\bar b_LH_1^*t_R
\right)+{\rm h.c.}
\label{eq:top-yukawa-expanded}
\end{equation}
First we consider the top-only axial-current coupling (the $SU(2)_L$
invariant requires the same coupling to $b_L$)
\begin{equation}
{\cal L}_{\phi t}
=\frac{\partial_\mu\phi}{f}
\left(
\bar t_R\gamma^\mu t_R
-\bar Q_L\gamma^\mu Q_L
\right).
\label{eq:top-only-current}
\end{equation}

We have four different three-body decay channels of $\phi$ related to top Yukawa couplings. For
\begin{equation}
\phi(p)\to H(k_2)+q_R(k_1)+\bar t_L(k_3),
\end{equation}
the matrix element is
\begin{equation}
{\cal M}_q
=-2i\frac{y_t}{f}\,
\bar u(k_1)P_Lv(k_3).
\label{eq:generic-yukawa-amplitude}
\end{equation}
Since the final-state chirality is already fixed by the projectors, the spin
sum gives
\begin{equation}
\sum_{\rm spins}|{\cal M}_q|^2
=4\frac{|y_t|^2}{f^2}
{\rm Tr}\!\left[
\slashed k_1P_L\slashed k_3P_R
\right]
=4\frac{|y_t|^2}{f^2}s_{13},
\qquad
s_{13}\equiv(k_1+k_3)^2 .
\label{eq:generic-spin-sum}
\end{equation}
With all final-state masses set to zero, the three-body phase space gives
\begin{equation}
d\Gamma_q
=\frac{1}{256\pi^3m_\phi^3}
\sum_{\rm spins}|{\cal M}_q|^2\,ds_{12}\,ds_{23},
\end{equation}
with
\begin{equation}
0\le s_{12}\le m_\phi^2,\qquad
0\le s_{23}\le m_\phi^2-s_{12}.
\end{equation}
Hence
\begin{equation}
\int ds_{12}\,ds_{23}\,s_{13}
=\frac{m_\phi^6}{6}.
\end{equation}
Therefore
\begin{equation}
\Gamma_q
=\frac{N_c|y_t|^2}{384\pi^3}
\frac{m_\phi^3}{f^2},
\label{eq:generic-single-color-width}
\end{equation}
with an overall color factor $N_c=3$. Combining all the four channels, the three-body width is
\begin{equation}
\Gamma_{\rm three-body}
=\frac{N_c|y_t|^2}{96\pi^3}
\frac{m_\phi^3}{f^2}
=\frac{|y_t|^2}{32\pi^3}
\frac{m_\phi^3}{f^2},
\qquad N_c=3 .
\label{eq:su2-total-width}
\end{equation}
which reproduces contact vertex three body decay results in~\eqref{eq:app-contact-width}. 

\subsection{3 body decay to fermion pair plus gauge boson}
\label{sec:Gauge3body}

\begin{figure}[htp!]
\centering
\begin{minipage}[c]{0.46\textwidth}
\centering
\begin{tikzpicture}[baseline=(current bounding box.center)]
\begin{feynman}
\vertex (phi) at (-1.65,0) {$\phi$};
\vertex[dot] (current) at (0,0) {};
\vertex (tb) at (1.75,-0.85) {$\bar \psi$};
\vertex[dot] (gauge) at (1.20,0.65) {};
\vertex (q) at (2.65,0.65) {$\psi$};
\vertex (g) at (1.20,1.65) {$g$};
\diagram*{
  (phi) -- [scalar] (current),
  (tb) -- [fermion] (current),
  (current) -- [fermion, edge label'=\(\psi\)] (gauge),
  (gauge) -- [fermion] (q),
  (gauge) -- [photon] (g)
};
\end{feynman}
\end{tikzpicture}
\par\vspace{-0.4em}\textbf{(a)}
\end{minipage}
\begin{minipage}[c]{0.46\textwidth}
\centering
\begin{tikzpicture}[baseline=(current bounding box.center)]
\begin{feynman}
\vertex (phi) at (-1.65,0) {$\phi$};
\vertex[dot] (current) at (0,0) {};
\vertex (q) at (1.75,0.85) {$\psi$};
\vertex[dot] (gauge) at (1.20,-0.65) {};
\vertex (tb) at (2.65,-0.65) {$\bar \psi$};
\vertex (g) at (1.20,-1.65) {$g$};
\diagram*{
  (phi) -- [scalar] (current),
  (gauge) -- [fermion, edge label=\(\psi\)] (current),
  (current) -- [fermion] (q),
  (tb) -- [fermion] (gauge),
  (gauge) -- [photon] (g)
};
\end{feynman}
\end{tikzpicture}
\par\vspace{-0.4em}\textbf{(b)}
\end{minipage}\hfill
\caption{The diagrams for
$\phi\to g\psi\bar{\psi}$ at original basis with $g$ corresponding to an arbitrary gauge boson.}
\label{fig:app-three-body-original-basis-gauge-boson}
\end{figure}
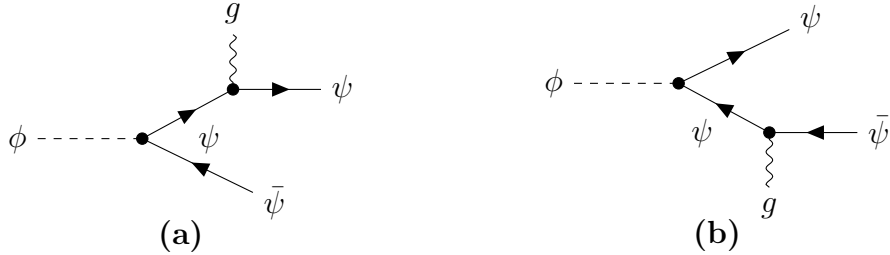

For completeness we here mention the scalar 3 body decay into the fermion pair plus gauge boson, assuming interaction of the form (\ref{delphi_psi5psi}).
We verify that the decay amplitude to fermion pair plus gauge boson is mass suppressed and vanishes in the massless fermion limit. Let us denote the inflaton momentum by $k$, the outgoing fermion momentum by $p$, outgoing antifermion momentum by $p^{\prime}$ and gauge field momentum by $q$. We assume massless gauge bosons here. We have two diagrams, one with the gauge boson emitted from the fermion and one with gauge boson emitted from the antifermion, as shown in Fig.~\ref{fig:app-three-body-original-basis-gauge-boson}.

The first diagram gives us
\begin{equation}
    \mathcal M\propto\bar{u}(p)\gamma^{\mu}\frac{\slashed{p}+\slashed{q}+m}{2p\cdot q}\slashed{k}\gamma_5 v(p^{\prime})\epsilon_\mu^*,
\end{equation}
where $m$ is the fermion mass, $\epsilon_{\mu}$ is the gauge field polarization vector and we suppress all the indices and structures corresponding to the gauge group. To see that this contains a term proportional to $m$ we write
\begin{equation}
    \slashed{k}\gamma_5 v(p^\prime)=(\slashed{p}+\slashed{q})\gamma_5v(p^{\prime})+\slashed{p}^{\prime}\gamma_5v(p^{\prime})=(\slashed{p}+\slashed{q}+m)\gamma_5v(p^{\prime}),
\end{equation}
After expanding $(\slashed{p}+\slashed{q}+m)^2$ we get
\begin{equation}
    \mathcal M\propto\bar{u}(p)\gamma^\mu\gamma_5v(p^{\prime})\epsilon_\mu^*+\frac{m}{p\cdot q}\bar{u}(p)\gamma^\mu(\slashed{p}+\slashed{q}+m)\gamma_5v(p^{\prime})\epsilon_\mu^*.
\end{equation}
For the antifermion case we have
\begin{equation}
        \mathcal M\propto\bar{u}(p)\slashed{k}\gamma_5\frac{-\slashed{p}^{\prime}-\slashed{q}+m}{2p^{\prime}\cdot q} \gamma^{\mu} v(p^{\prime})\epsilon_\mu^*.
\end{equation}
Commuting the $\gamma_5$ through we get 
\begin{equation}
    \mathcal M\propto-\bar{u}(p)\slashed{k}\frac{+\slashed{p}^{\prime}+\slashed{q}+m}{2p^{\prime}\cdot q} \gamma^{\mu}\gamma_5 v(p^{\prime})\epsilon_\mu^*
\end{equation}
and by analogical manipulation to the first diagram we get the same expression with $p$ replaced by $p^{\prime}$, but with an opposite sign. The terms not proportional to $m$ cancel out.
This finding is consistent with the argument given in Appendix~\ref{app:torsion_coupling}.

\subsection{Loop channels and current selection rules}

The same basis also clarifies several channels that vanish at the order
considered here. 
First note that the interaction of the form (\ref{eq:canonical-torsion-coupling}) indicates that the torsion scalar $\phi$ is CP odd.
A two-Higgs final state is CP even, so a CP-odd
torsion cannot decay into it through a one-loop diagram with real
Yukawa couplings.  In a direct trace calculation, every potentially nonzero
CP-odd term requires four gamma matrices in addition to $\gamma_5$; the
two-Higgs amplitude does not supply the required Lorentz structure.

\begin{figure}[t]
\centering
\begin{minipage}[c]{0.31\textwidth}
\centering
\begin{tikzpicture}[baseline=(current bounding box.center),scale=0.9]
\begin{feynman}
\vertex[dot] (c) at (-0.75,0) {};
\vertex[dot] (h) at (0.85,0.55) {};
\vertex[dot] (v) at (0.85,-0.80) {};
\vertex (phi) at (-1.85,0) {$\phi$};
\vertex (hi) at (1.65,1.45) {$H_i^\dagger$};
\vertex (hj) at (2.25,0.55) {$H_j$};
\vertex (vv) at (2.25,-0.80) {$V^A$};
\diagram*{
  (phi) -- [scalar] (c),
  (c) -- [scalar] (hi),
  (h) -- [fermion] (c),
  (v) -- [fermion] (h),
  (c) -- [fermion] (v),
  (h) -- [scalar] (hj),
  (v) -- [boson] (vv)
};
\node at (0.12,-0.10) {$Q_L$};
\node at (1.23,-0.12) {$t_R$};
\node at (0.02,-0.76) {$t_R$};
\node at (0.25,-1.35) {\textbf{(a)}};
\end{feynman}
\end{tikzpicture}
\end{minipage}\hfill
\begin{minipage}[c]{0.31\textwidth}
\centering
\begin{tikzpicture}[baseline=(current bounding box.center),scale=0.9]
\begin{feynman}
\vertex[dot] (c) at (-0.75,0) {};
\vertex[dot] (h) at (0.85,0.55) {};
\vertex[dot] (v) at (0.85,-0.80) {};
\vertex (phi) at (-1.85,0) {$\phi$};
\vertex (hi) at (1.65,1.45) {$H_i^\dagger$};
\vertex (hj) at (2.25,0.55) {$H_j$};
\vertex (vv) at (2.25,-0.80) {$V^A$};
\diagram*{
  (phi) -- [scalar] (c),
  (c) -- [scalar] (hi),
  (c) -- [fermion] (h),
  (h) -- [fermion] (v),
  (v) -- [fermion] (c),
  (h) -- [scalar] (hj),
  (v) -- [boson] (vv)
};
\node at (0.12,-0.10) {$t_R$};
\node at (1.23,-0.12) {$Q_L$};
\node at (0.02,-0.76) {$Q_L$};
\node at (0.25,-1.35) {\textbf{(b)}};
\end{feynman}
\end{tikzpicture}
\end{minipage}\hfill
\begin{minipage}[c]{0.31\textwidth}
\centering
\begin{tikzpicture}[baseline=(current bounding box.center),scale=0.9]
\begin{feynman}
\vertex (phi) at (-1.35,0) {$\phi$};
\vertex[dot] (a) at (0,0) {};
\vertex (vout) at (1.25,1.0) {$V^A$};
\vertex[dot] (s) at (1.25,-0.65) {};
\vertex (h1) at (2.65,-0.1) {$H$};
\vertex (h2) at (2.65,-1.2) {$H^\dagger$};
\diagram*{
  (phi) -- [scalar] (a),
  (a) -- [boson] (vout),
  (a) -- [boson,edge label'={$V^{A*}$}] (s),
  (s) -- [scalar] (h1),
  (s) -- [scalar] (h2)
};
\node at (0.55,-1.55) {\textbf{(c)}};
\end{feynman}
\end{tikzpicture}
\end{minipage}
\caption{Electroweak gauge-boson decay in the current-divergence basis.  In panels (a) and (b), the
leftmost loop vertex is the
$\phi\,\bar t_R\widetilde H^\dagger Q_L$ contact interaction.  The diagrams with the two external Higgs legs interchanged are
implicit.
Panel (c) is the anomaly-mediated contribution with one off-shell gauge boson.}
\label{fig:app-hhv-current-diagrams}
\end{figure}
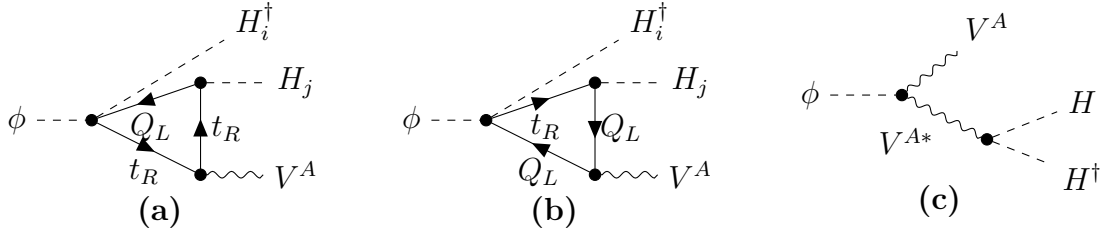

For the ordering
$\phi(p)\to H_i^\dagger(k_1)+H_j(k_2)+V^A(q)$, attaching an electroweak gauge
boson leads to the two triangle diagrams shown
in Figs.~\ref{fig:app-hhv-current-diagrams} (a) and
\ref{fig:app-hhv-current-diagrams} (b).  For the diagram with gauge emission
from the $t_R$ line, the CP-odd numerator can be reduced to
\begin{align}
N_R^\alpha\big|_{\rm odd}
&\propto\epsilon^{\alpha\mu\nu\rho}
(\ell+k_1+q)_\mu(\ell+k_1)_\nu\ell_\rho
\notag\\
&=\epsilon^{\alpha\mu\nu\rho}q_\mu k_{1\nu}\ell_\rho.
\label{eq:app-hhv-odd-numerator}
\end{align}
After Feynman parametrization, the loop shift has the form
$\ell=L-ak_1-bq$.  The term linear in $L$ integrates to zero, while the
remaining terms contain
$\epsilon^{\alpha\mu\nu\rho}q_\mu k_{1\nu}(ak_1+bq)_\rho=0$.
The $Q_L$-emission diagram and the charge-conjugate pair vanish by the same
argument.  A nonzero $\phi\to HH^\dagger V$ amplitude can arise from the
anomaly operator through an off-shell gauge boson, as shown in
Fig.~\ref{fig:app-hhv-current-diagrams} (c).  This contribution carries an
additional gauge interaction and has no top-Yukawa enhancement, so it is
subleading to the two-gauge-boson and Yukawa-assisted three-body widths kept in
the reheating analysis.

\subsection{Comment on the case of vector current interaction}

In the most part of this paper we assumed the scalar derivative interaction with axial current as described by (\ref{delphi_psi5psi}), since it naturally arises from the minimal torsion-matter coupling. 
In general, however, torsion or distortion fields can have a coupling with a vector current as\footnote{
    This type of interaction is often introduced in the context of spontaneous baryogenesis~\cite{Cohen:1987vi,Cohen:1988kt}. Its relation to the SU(2) gauge anomaly in the axion-like model has been explored in Ref.~\cite{Kusenko:2014uta}. Note that this term is consistent with CP if $\phi$ is CP-odd, but breaks C and P individually.
}
\begin{align}
	\mathcal L = \frac{\partial_\mu \phi}{f}\sum_i\overline\psi_i\gamma_\mu\psi_i.
    \label{delphi_psipsi}
\end{align}
The vector rotation of fermions, $\psi_{L_i} \to e^{i\theta_L}\psi_{Li}$ and $\psi_{R_i} \to e^{i\theta_R}\psi_{Ri}$ with $\theta_L=\theta_R=-\phi/f$, can remove this coupling.
The remaining action, including mass terms or gauge/Yukawa interaction terms, are invariant under this rotation.\footnote{
    Majorana mass terms are not invariant. We do not consider them here since we are mainly interested in the Standard Model fermions.
}
Thus the 2-body decay $\phi\to\overline\psi\psi$, as well as many body decays, vanish independently of the fermion mass.
The only non-zero contribution arises from the anomaly under chiral gauge groups. 
Let us suppose that some of $\psi_{Li}$ and $\psi_{Ri}$ are fundamental representations of chiral gauge group SU($N$). Then we have the following term after the vector rotation:
\begin{align}
	\mathcal L = \sum_i(C_{Ri}-C_{Li})\frac{\alpha_g}{2\pi}\frac{\phi}{f} W^a_{\mu\nu}\widetilde W^{\mu\nu a},
\end{align}
where $C_{Li}$ or $C_{Ri}$ is $1/2$ if the corresponding left- or right-handed fermions are fundamental representations, while they are zero otherwise. The decay rate to the gauge boson pair is
\begin{align}
	\Gamma(\phi\to WW) \simeq \frac{(N^2-1)\alpha_g^2}{16\pi^3} \left[\sum_{i}\left(C_{Ri}-C_{Li}\right) \right]^2 \frac{m_\phi^3}{f^2}.
\end{align}
In the Standard Model gauge groups, the SU(2) gauge group is chiral and $C_{Ri}=0$ for all fermions.

\bibliography{references}

\end{document}